\documentclass[aps,prl,twocolumn,superscriptaddress]{revtex4-2}

\usepackage{graphicx}
\usepackage{dcolumn}
\usepackage{bm}
\usepackage{hyperref}
\hypersetup{
    colorlinks=true,
    linkcolor=blue,
    filecolor=magenta,      
    urlcolor=red,
    citecolor=blue,
}
\usepackage{todonotes}
\usepackage{color}
\usepackage{glossaries}
\usepackage[version=4]{mhchem}
\usepackage[
width=182mm, columnsep=6mm, height=9in, includefoot,
]{geometry}

\makeatletter
\renewcommand*{\fnum@figure}{{\normalfont\bfseries \figurename~\thefigure}}  
\makeatother

\renewcommand{\figurename}{Fig.}

\newacronym{2D}{2D}{two-dimensional}
\newacronym{vdW}{vdW}{van der Waals}
\newacronym{RIXS}{RIXS}{resonant inelastic x-ray scattering}
\newacronym{XAS}{XAS}{x-ray absorption spectrum}
\newacronym{IXS}{IXS}{inelastic x-ray scattering}
\newacronym{EELS}{EELS}{electron energy loss spectroscopy}
\newacronym{ED}{ED}{exact diagonalization}
\newacronym{AIM}{AIM}{Anderson impurity model}
\newacronym{FWHM}{FWHM}{full-width at half-maximum}
\newacronym{DHO}{DHO}{damped harmonic oscillator}
\begin{document}

\title{Magnetically propagating Hund's exciton \\in van der Waals antiferromagnet NiPS$_3$}

\author{W. He}
\email{whe1@bnl.gov}
\author{Y. Shen}
\affiliation{Department of Condensed Matter Physics and Materials Science, Brookhaven National Laboratory, Upton, New York 11973, USA\looseness=-1}

\author{K. Wohlfeld}
\affiliation{Institute of Theoretical Physics, Faculty of Physics, University of Warsaw, Pasteura 5, PL-02093 Warsaw, Poland\looseness=-1}

\author{J. Sears}
\affiliation{Department of Condensed Matter Physics and Materials Science, Brookhaven National Laboratory, Upton, New York 11973, USA\looseness=-1}

\author{J. Li}
\author{J. Pelliciari}
\affiliation{National Synchrotron Light Source II, Brookhaven National Laboratory, Upton, New York 11973, USA\looseness=-1}

\author{M. Walicki}
\affiliation{Institute of Theoretical Physics, Faculty of Physics, University of Warsaw, Pasteura 5, PL-02093 Warsaw, Poland\looseness=-1}

\author{S.~Johnston}
\affiliation{Department of Physics and Astronomy, The University of Tennessee, Knoxville, Tennessee 37966, USA\looseness=-1}
\affiliation{Institute of Advanced Materials and Manufacturing, The University of Tennessee, Knoxville, Tennessee 37996, USA\looseness=-1}

\author{E. Baldini}
\affiliation{Department of Physics, The University of Texas at Austin, Austin, Texas, USA, 78712}

\author{V. Bisogni}
\affiliation{National Synchrotron Light Source II, Brookhaven National Laboratory, Upton, New York 11973, USA\looseness=-1}

\author{M. Mitrano}
\affiliation{Department of Physics, Harvard University, Cambridge, Massachusetts 02138, USA\looseness=-1}

\author{M. P. M. Dean}
\email{mdean@bnl.gov}
\affiliation{Department of Condensed Matter Physics and Materials Science, Brookhaven National Laboratory, Upton, New York 11973, USA\looseness=-1}

\def\mathbi#1{\ensuremath{\textbf{\em #1}}}
\def\Q{\ensuremath{\mathbi{Q}}}
\newcommand{\angstrom}{\mbox{\normalfont\AA}}
\date{\today}

\begin{abstract}
Magnetic \gls*{vdW} materials have opened new frontiers for realizing novel many-body phenomena. Recently \ce{NiPS3} has received intense interest since it hosts an excitonic quasiparticle whose properties appear to be intimately linked to the magnetic state of the lattice. Despite extensive studies, the electronic character, mobility, and magnetic interactions of the exciton remain unresolved. Here we address these issues by measuring \ce{NiPS3} with ultra-high energy resolution \gls*{RIXS}. We find that Hund's exchange interactions are primarily responsible for the energy of formation of the exciton. Measuring the dispersion of the Hund's exciton reveals that it propagates in a way that is analogous to a double-magnon. We trace this unique behavior to fundamental similarities between the \ce{NiPS3} exciton hopping and spin exchange processes, underlining the unique magnetic characteristics of this novel quasiparticle. 
\end{abstract}

\maketitle

\section{Introduction \label{sec:introduction}}
\Gls*{2D} \gls*{vdW} materials provide an ideal platform for combining strong electronic correlations, low-dimensional magnetism, and weak dielectric screening to realize novel electronic quasiparticles and functionality \cite{Burch2018review, Gong2019two, Wang2022review}. Recent years have seen the identification of excitons in a host of closely related \gls*{vdW} compounds such as \ce{NiPS3}, \ce{CrSBr}, \ce{NiI2}, and \ce{MnPS3} \cite{Kang2020Coherent, Son2022Multiferroic, Gnatchenko2011exciton, Wang2023excitonmagnon}. Within this family, the \ce{NiPS3} exciton exhibits several fascinating properties, including  strong interactions between the exciton lifetime and magnetic order \cite{Kang2020Coherent}, thickness-dependent properties in the few-layer-limit \cite{Hwangbo2021Highly, Wang2022Electronic, DiScala2024elucidating}, coupling between magnetism and exciton polarization \cite{Hwangbo2021Highly, Wang2021Spininduced, Afanasiev2021Controlling}, and unconventional exciton-driven metallic behavior \cite{Belvin2021Excitondriven}. These observations suggest that excitons in magnetic \gls*{vdW} materials such as \ce{NiPS3} might have fundamentally different character from other types of exciton, such as the Frenkel, Wannier, and Hubbard varieties. Frenkel and Wannier excitons form via Coulomb interactions between electrons and holes in different Bloch states and propagate according to the detailed form of the band-structure and electron-hole attraction \cite{Wang2018Colloquium}. Hubbard excitons, on the other hand, form from strongly correlated many-body states and their propagation is expected to involve the scattering of spin waves \cite{Kim2014Excitonic, Huang2023spin}.

\begin{figure*}[!ht]
\includegraphics{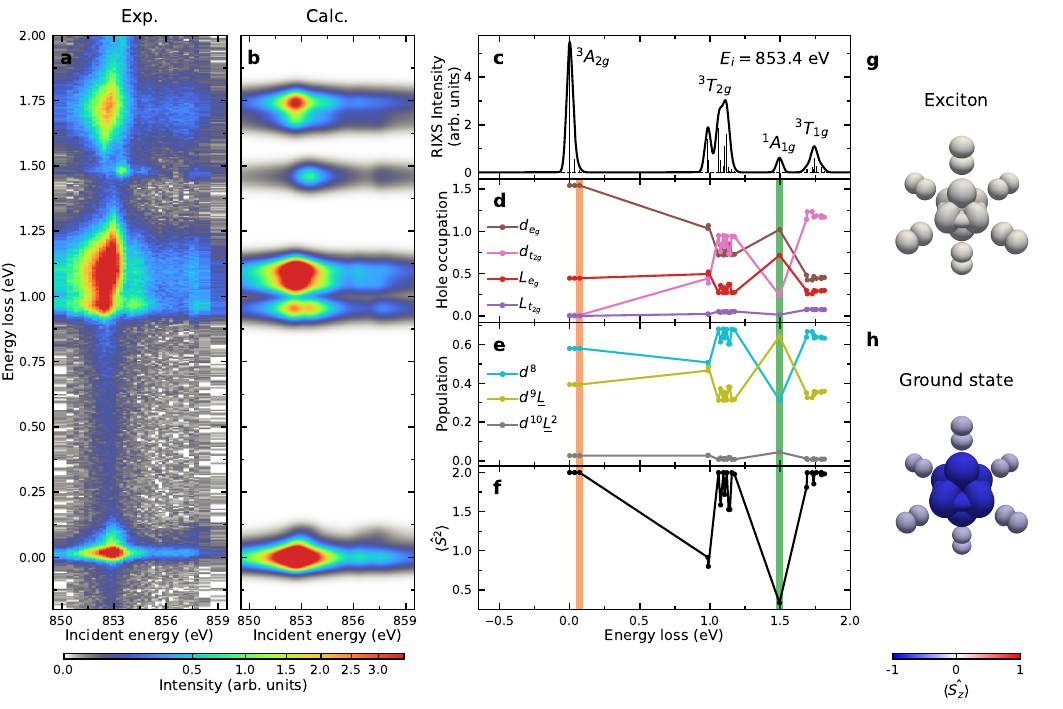}
\caption{\textbf{Electronic character of the NiPS$_3$ exciton}. \textbf{a}, \gls*{RIXS} intensity map as a function of incident photon energy through the Ni $L_3$ resonance. The exciton is visible at an energy loss of $1.47$~eV and reaches a maximum intensity at an incident energy of $853.4$~eV. These data were taken at 40~K with $\pi$-polarized x-rays incident on the sample at $\theta = 22.6^\circ$ and scattered to $2\Theta = 150^\circ$. \textbf{b}, \gls*{RIXS} calculations for \ce{NiPS3} that capture the energy and resonant profile of the $dd$-transitions and exciton in the material. \textbf{c}, Calculated unbroadened \gls*{RIXS} intensity (vertical lines) and broadened \gls*{RIXS} spectra (solid curve) at the main resonant incident energy of the exciton peak (i.e., $E_i = 853.4$~eV). \textbf{d}--\textbf{f}, Description of the ground and excited states in \ce{NiPS3}. \textbf{d} shows the hole occupations of Ni $3d$ (denoted by $d$) and ligand (denoted by $L$) orbitals. \textbf{e} displays probabilities of having $d^8$, $d^9\underline{L}$, and $d^{10}\underline{L}^2$ configurations. \textbf{f} gives the expectation value of the total spin operator squared $\langle \hat{S}^2 \rangle$. The orange (green) vertical lines in \textbf{d}--\textbf{f} indicate the energy for the double-magnons (excitons). \textbf{g},\textbf{h}, Wavefunction illustrations extracted from \textbf{b} for \textbf{g} the exciton and \textbf{h} the ground state. The size of each orbital ($3d$ for the central Ni site and $3p$ for the six neighboring S sites) is proportional to its hole occupation.  The color represents the expectation value of the spin operator along the $z$ axis $\langle \hat{S}_z \rangle$, again calculated separately for the Ni and S states. Therefore, the change in spin state and the partial transfer of holes involved in the exciton transition is encoded in the change in color and size of orbitals, respectively. We represent the ground state by only the down-spin configuration, omitting the up-spin and spin-zero elements of the triplet.}
\label{fig:energy_map}
\end{figure*}

In previous works, the \ce{NiPS3} exciton has been described as a ``Zhang-Rice'' mode \cite{Kang2020Coherent}. This terminology derives from studies of cuprate superconductors, and refers to a specific form of hybridized wavefunctions that have one hole on the transition metal (Ni) site and one hole on the ligand (S) site and describes the ``Zhang-Rice exciton'' as a transition from a high-spin triplet to a low-spin singlet \cite{Zhang1988Effective}.  However, the way the exciton changes with applied magnetic field has been argued to be incompatible with this picture \cite{Jana2023magnon}. A Zhang-Rice scenario also does not address why the exciton has such a narrow linewidth \cite{Wang2021Spininduced, Dirnberger2022Spincorrelated, Hwangbo2021Highly}. The unsettled and possibly unconventional exciton electronic character suggests that the exciton may also propagate in an exotic manner different to regular excitons, but, to date, this has never been measured. Here, we use ultra-high energy resolution \gls*{RIXS} to directly detect the \ce{NiPS3} exciton momentum dispersion and discover it propagates magnetically in a similar way to the double-magnon excitation. Through detailed analysis of the exciton wavefunction, we further reveal the different interactions involved in its formation and establish that its primary character is that of a Hund's exciton, distinct from the Zhang-Rice and other scenarios. 

\section{Results \label{sec:results}}
\subsection{Electronic character of the exciton\label{sec:results_Emap}}

We start by measuring the incident energy dependence of the Ni $L_3$-edge \gls*{RIXS} spectrum of \ce{NiPS3} to identify the different spectral features present (see Fig.~\ref{fig:energy_map}a and Methods). The most intense peaks centered around $1.0$, $1.1$, and $1.7$~eV energy loss are $dd$ excitations in which electrons transition between different Ni $3d$ orbitals. A remarkably sharp (almost resolution limited) peak is apparent at an energy loss matching the known energy of the exciton at $1.47$~eV. This excitation resonates strongly at $853.4$~eV and is well separated from other $dd$ and the higher energy charge-transfer excitations. We identify this feature as the \ce{NiPS3} exciton, consistent with previous reports \cite{Kang2020Coherent}. 

\begin{figure*}
\includegraphics{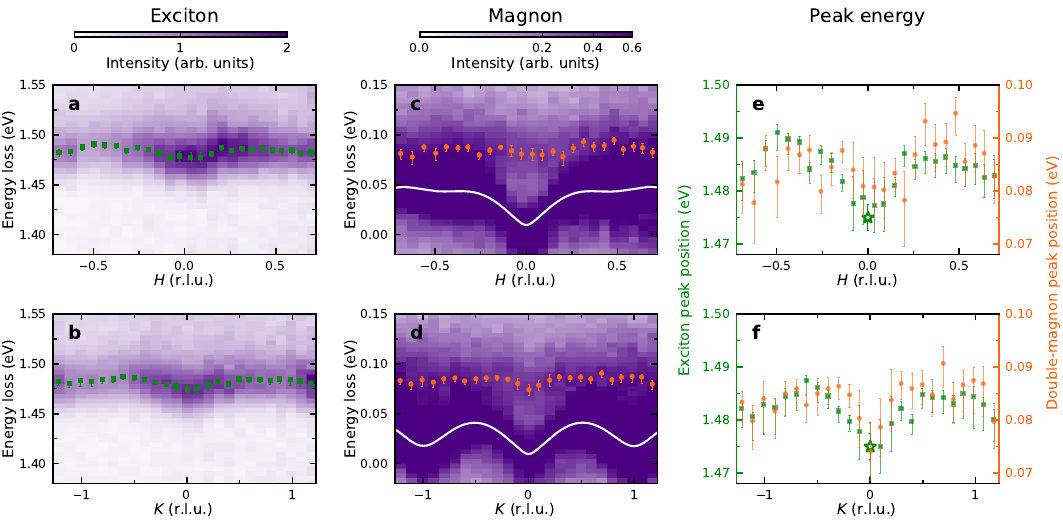}
\caption{\textbf{Low temperature exciton dispersion and comparison with double-magnons}. \textbf{a},\textbf{b}, \gls*{RIXS} intensity maps as a function of the $H$ and $K$ in-plane momentum transfer, respectively, with an energy window chosen to isolate the exciton dispersion. The overlaid green squares mark the peak positions of the exciton. \textbf{c} and \textbf{d} show the low energy dispersion at equivalent momenta with the observed inelastic feature, including magnons (white lines) and double-magnons (orange circles). Panels \textbf{e} and \textbf{f}, show that both the exciton and double-magnon have similar dispersion with an energy offset of $\sim 1.4$~eV. All the measurements were taken at $T=40$~K using $\pi$-polarized incident x-rays at an incident energy of $853.4$~eV corresponding to the exciton resonance. The asterisks in panels \textbf{e} and \textbf{f} denote the reported exciton energy from optical measurements \cite{Kang2020Coherent, Hwangbo2021Highly, Wang2021Spininduced} with error bars from our instrument energy calibration (one standard deviation). All other error bars are 1-$\sigma$ confidence intervals evaluated from the fitting as explained in the Methods. Detailed linecuts showing the fitting are provided in Supplementary Figs.~3--5.}
\label{fig:dispersion_LT}
\end{figure*}

To facilitate our understanding of the exciton and its interplay with magnetism, we constructed an effective \ce{NiS6} cluster model representing \ce{NiPS3} (See Supplementary Note 1A). Our model includes Coulomb repulsion, Hund's coupling, crystal field, and Ni-S and S-S hopping. As explained in the Methods section, the rich spectrum, including the detailed splitting of the two $dd$-excitations at 1.0 and 1.1 eV allows us to obtain a well-constrained model Hamiltonian for \ce{NiPS3} (see Supplementary Note 1B and Supplementary Fig.~2).

To better understand the nature of the exciton, we plot several expectation values describing the \ce{NiPS3} wavefunction in Fig.~\ref{fig:energy_map}d--f, which reveal that \ce{NiPS3} has dominant Hund's rather than Zhang-Rice character. The plotted expectation values include the hole occupations of the Ni $3d$ and ligand orbitals and the weights of the $d^8$, $d^9\underline{L}$, and $d^{10}\underline{L}^2$ configurations ($\underline{L}$ stands for a ligand hole) that make up each state. We also calculate the expectation values for the total spin operator squared $\hat{S}^2$, which, for two holes, has a maximum value of $S(S+1)=2$. The ground state is close to this pure high spin state; while the spin is strongly reduced in the exciton. We therefore confirm that the exciton is dominantly a triplet-singlet excitation. In the Zhang-Rice scenario, the leading character of the ground state would be $d^9\underline{L}$. The dominant component is, in fact, $d^8$ revealing that the state has dominant Hund's character.

The ground state and the exciton wavefunctions obtained from our model are illustrated in Fig.~\ref{fig:energy_map}g,h and are described in detail in Supplementary Note 2. We see that substantial charge redistribution occurs during exciton formation, which is partly crystal field and partly a Ni-S charge transfer in nature (see Fig.~\ref{fig:energy_map}d,e). As explained later in the discussion section, the fact that the ground states have dominant Ni character (rather than dominant Zhang-Rice character) plays a leading role in the energy for exciton formation. 

\subsection{Exciton dispersion \label{sec:results_dispersion}}

Having clarified the character of the exciton, we study its propagation by tuning the incident x-ray energy to the exciton resonance at $853.4$~eV and mapping out the in-plane dispersion with high energy resolution (Fig.~\ref{fig:dispersion_LT}a and b). The two high-symmetry in-plane reciprocal space directions both exhibit a small upward dispersion away from the Brillouin zone center with similar bandwidths of $\sim15$~meV. This non-zero dispersion suggests that the exciton excites low-energy quasiparticles as it propagates through the lattice. We consequently, mapped out the low energy excitations in Fig.~\ref{fig:dispersion_LT}c,d. The strongest feature is the magnon, which was found to be consistent with the prediction based on prior inelastic neutron scattering measurements \cite{Wildes2022Magnetic, Scheie2023Spin} (the white line, see Methods). Intriguingly, we also found another broad low-energy dispersive feature at an energy scale roughly twice as large as the magnon peak (the orange dots in Fig.~\ref{fig:dispersion_LT}c and d). As we justify in detail later, the observed low-energy feature, in fact, corresponds to ``double-magnon'' excitations, which is a process in which two spins are flipped on each site making up the excitation to create a pair of magnons with the same spin. By fitting the exciton and double-magnon energy, we see that these two excitations show similar dispersion despite their drastically different energy scales (Fig.~\ref{fig:dispersion_LT}e and f), indicating that the exciton propagates in a way that is similar to the double-magnon. Since the dispersive effects are subtle, we confirmed the calibration of the spectra by verifying that the magnon energy exactly reproduces the dispersion obtained in prior inelastic neutron scattering experiments \cite{Wildes2022Magnetic, Scheie2023Spin}. We similarly confirmed that the Brillouin zone center exciton energy is consistent with values from optical measurements \cite{Kang2020Coherent, Hwangbo2021Highly, Wang2021Spininduced}.

\subsection{Temperature dependence \label{sec:results_tdep}}
\begin{figure}
\includegraphics{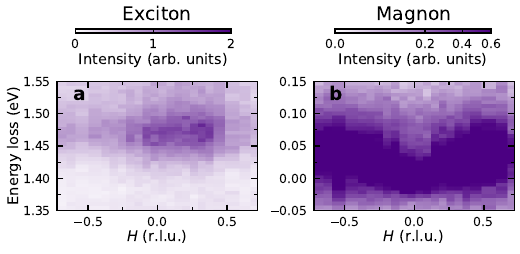}
\caption{\textbf{High temperature exciton dispersion and comparison with double-magnons above the N\'{e}el transition}. The \gls*{RIXS} intensity maps with energy window chosen to isolate the excitons (\textbf{a}) and magnons/double-magnons (\textbf{b}) as a function of in-plane momentum transfer $H$ measured in the ($H0L$) scattering plane. All the measurements were taken at $T=190$~K with linear horizontal $\pi$ polarization of the incident x-rays at the resonant energy of $853.4$~eV for excitons. The same data are provided as linecuts in Supplementary Figs.~9 and 10.}
\label{fig:dispersion_HT}
\end{figure}

To substantiate the identity of the low energy excitations we observe in \ce{NiPS3}, we need to consider the \gls*{RIXS} cross-section. Since \gls*{RIXS} is a photon-in photon-out scattering process with each photon carrying one unit of angular momentum, it can couple to processes involving either zero, one, or two spin flips \cite{Groot1998local, Ament2011resonant, Ghiringhelli2009, Fabbris2017doping, Nag2020many}. While the spin-flip processes are necessarily magnetic, the zero spin flip process can either correspond to a phonon or to a so called ``bi-magnon'' in which two magnons with opposite spins are created on neighboring sites to make an excitation with a net spin of zero. The most straightforward way to distinguish magnetic and non-magnetic excitations is to measure the dispersion above the N\'{e}el order temperature of $T_{\rm N}=159$~K, as presented in Fig.~\ref{fig:dispersion_HT}. Above $T_{\rm N}$ the exciton remains visible, but it becomes weaker and more diffuse compared to the data at 40~K (see Fig.~\ref{fig:dispersion_HT}a and the linecuts in Supplementary Fig.~10). Consequently, no dispersion is detectable. The double-magnon peak that was observed at 40~K is replaced by a diffuse, over-damped tail of intensity, contrary to what would be expected for a phonon and corroborating its magnetic origin (see Fig.~\ref{fig:dispersion_HT}b and the linecuts in Supplementary Fig.~9). The residual intensity arises from short-range spin fluctuations, which are expected to persist well above $T_{\rm N}$ in quasi-\gls*{2D} magnets as long as the thermal energy scale is well below the energy scale of the magnetic interactions \cite{Ronnow2001spin}. We also note that optical phonons in the $70$--$100$~meV energy range in \ce{NiPS3} are known to be minimally-dispersive \cite{Bernasconi1988Lattice}, which again suggests a magnetic origin. 

\subsection{Identifying double-magnons through their resonant profile \label{sec:results_resonance}}

\begin{figure}
\includegraphics{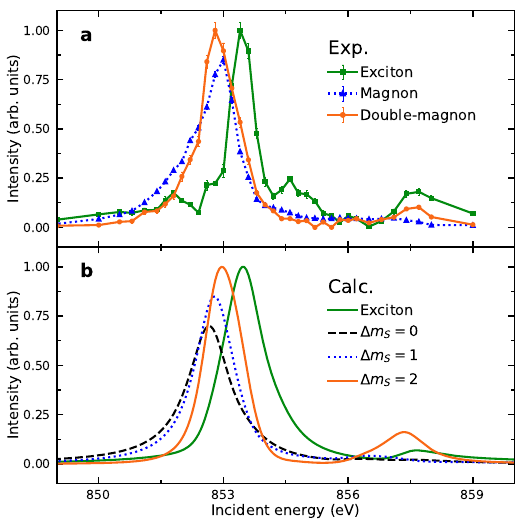}
\caption{\textbf{Resonance behavior of magnons, double-magnons and excitons}. \textbf{a}, The measured \gls*{RIXS} spectral weights of the magnon, double-magnon and exciton extracted by fitting experimental \gls*{RIXS} spectrum for each incident energy. Error bars represent one standard deviation. Data were taken at 40~K at a scattering angle of $2\Theta = 150^\circ$ and an incident angle of $\theta = 22.6^\circ$. \textbf{b}, The calculated \gls*{RIXS} spectral weights of the exciton and low-energy zero-, one-, and two-spin-flip transitions ($\Delta m_s=0, 1, 2$) as a function of incident energy. The curves in both panels are scaled for clarity. }
\label{fig:resonance}
\end{figure}

Having established a magnetic origin for the dispersion, we can use the energy-dependent resonant profile of \gls*{RIXS} to distinguish different magnetic processes and substantiate our assignment of the low-energy feature as the double-magnon. Figure~\ref{fig:resonance}a plots the energy dependence of the measured double-magnon feature alongside the magnon and the exciton. We compare this with calculations of the \gls*{RIXS} cross-section based on our model, which includes processes involving zero, one, and two spin flips denoted by $\Delta m_S=0$, $1$, and $2$, respectively, as well as coupling to the exciton. $\Delta m_S=0$ reflects the cross section for either elastic scattering or a bi-magnon, $\Delta m_S=1$ corresponds to a magnon, and $\Delta m_S=2$ corresponds to the double-magnon. Notably, since the double-magnon involves an exchange of two units of spin angular momentum, \gls*{RIXS} is especially suitable for detecting this process. The main resonance around 853~eV shows partial overlap between the double-magnon and bi-magnon resonance, so it does not clearly distinguish between them. However, the presence of the excitation at the satellite resonance at 857.7~eV is only compatible with a $\Delta m_S=2$ double-magnon process. We therefore suggest that the dominant character of the excitation around 80~meV is that of double-magnons substantiating our spectral assignment, although we cannot exclude a small sub-leading contribution. Our assignment is also supported by \gls*{RIXS} studies of NiO where the double-magnon was also found to be much more intense than the bi-magnon \cite{Groot1998local, Ghiringhelli2009, Betto2017three, Nag2020many}. The satellite resonance is generated by the exchange part of the core-valence Coulomb interaction on the \ce{Ni} site. This same interaction creates the double-magnon because the angular momentum state of the core-hole needs to vary in the intermediate state in order to allow two subsequent spin-flip processes to occur in the photon absorption and emission processes. The core-valence exchange interaction facilitates this change in the core-hole state by mixing of the core-hole and valence eigenstates. This conclusion is also borne out in studies of NiO where the same process occurs \cite{Groot1998local, Ghiringhelli2009, Betto2017three, Nag2020many}.

\begin{figure}
\includegraphics[width=1.0\columnwidth]{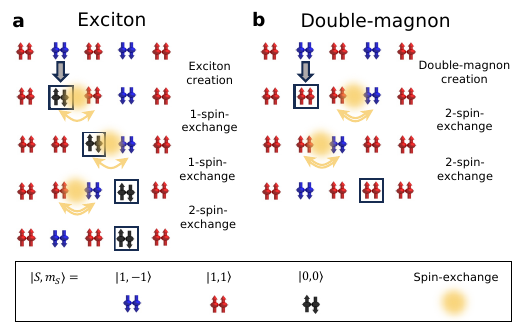}
\caption{\textbf{Illustration of exciton and double-magnon propagation based on perturbation theory}. The top row represents the antiferromagnetic background and subsequent rows show the time evolution of the state. \textbf{a}, After the singlet $\vert 0, 0 \rangle$ exciton forms (second row from the top) it exchanges spin with neighboring sites such that it moves while flipping spins and breaking magnetic bonds; free propagation to the next nearest neighbor site (bottom row) is possible after four spin exchanges, involving up to four magnons created in the intermediate state (middle rows). \textbf{b}, After the double-magnon excitation is created on the same site (second row from the top), it can freely move to the next nearest neighbor (bottom row) by four spin exchanges and exciting four magnons in the intermediate state (middle rows). The similarities between the propagation in \textbf{a} and \textbf{b} rationalize the experimentally observed similar dispersion relation of the exciton and double-magnon. These processes are mediated by the different spin-exchange interactions, with the third nearest neighbor exchange process playing the leading role. 
}
\label{fig:cartoon}
\end{figure}

\section{Discussion \label{sec:discussion}}
In this work we use high resolution \gls*{RIXS} to assess the formation and propagation of the excitonic state of \ce{NiPS3}. By combining \gls*{RIXS} and \gls*{ED} calculations, we reveal that the primary mechanism behind the exciton formation is the Hund's interaction. As illustrated in Fig.~\ref{fig:energy_map}d-h the exciton forms from a ground state with dominant $d^8$ character and involves significant charge transfer and crystal field changes. As such, the state we identify is quite different from prior descriptions of a Zhang-Rice exciton \cite{Kang2020Coherent}. As we discuss later, the distinction between these models is crucial as it corresponds to a different majority component of the wavefunction, different interactions playing the leading role in the exciton energy, and the possibility of realizing a model with physically reasonable parameters. These issues will be central to efforts to manipulate the exciton energy and cross section. We found that the difference in the prior identification of the exciton arises from using an under-constrained model. If one considers just the exciton energy and assumes that Hund's coupling can take any value, there are a range of different Hund’s interactions and charge transfer energy parameters that predict a 1.47~eV exciton. If one adds a further constraint that the $1.0$, $1.1$, and $1.7$~eV $dd$-excitations must be reproduced within an accuracy similar to their width, properly constrained solutions can be identified (see Supplementary Note 1F). Importantly, the solution found here also yields a physically reasonable value for Hund's coupling $J_H= 1.24$~eV corresponding to 87\% of the atomic value. This is relevant because the pure triplet and singlet Zhang-Rice components of the wavefunctions are energetically split by rather weak Ni-S exchange processes, so it is difficult to justify the $1.47$~eV energy scale of the exciton within a model with dominant Zhang-Rice character. In Ref.~\cite{Kang2020Coherent} an unphysically large Hund's coupling corresponding to 120\% of the atomic value was required.

Based on the wavefunction extraction performed in this study, we can determine which electronic interactions play the leading role in the exciton energy. To do this, we factorized the wavefunctions into the singlet and triplet components of the $d^8$, $d^9\underline{L}$, and $d^{10}\underline{L}^2$, configurations as described in Supplementary Note 2. We can then compute the contributions of Hund's, charge transfer, and crystal field to each of these components. We find that the primary contribution to the exciton energy comes from singlet-triplet splitting of the $d^8$ component of the wavefunction, which means that it is best thought of as a Hund's exciton since its energy of formation is mostly driven by Hund's exchange. Our exciton character derived here, also retains partial magnetic character coming from a sizable contribution of states beyond the three $d$-electron, Zhang-Rice, and ligand, singlet states. This means that the exciton is expected to vary with applied magnetic field compatible with recent observations \cite{Jana2023magnon}. It also implies that future efforts to realize similar symmetry excitons with different energies should target means to modify the on-site Ni Hund's exchange coupling and not the Ni-S exchange processes that would be the leading contribution to the energy of a Zhang-Rice exciton.

Our detection of exciton dispersion in \ce{NiPS3} proves that the exciton is an intrinsic propagating quasiparticle and excludes prior suggestions that the exciton might be a localized phenomenon associated with defects \cite{Kim2023Anisotropic}. The most common form of exciton propagation in weakly correlated transition metal chalcogenides involves excitations that are composed of bound pairs of specific Bloch states \cite{Hybertsen1986electron, Rohlfing2000}. The \ce{NiPS3} exciton is quite different from the more conventional excitons since it is bound by the local Hund's interactions described previously, rather than long-range Coulomb attraction. Recent calculations that work for many more conventional excitons, indeed fail to capture our measured exciton dispersion \cite{Lane2022ab}. We also note that the two holes in the Ni $e_g$ manifold represent what is in some sense the simplest way to realize a triplet-singlet excitation. Consequently, \ce{NiPS3} has relatively few excitations compared to other materials and the exciton is energetically well separated from other transitions, such that it interacts exclusively by magnons and not other types of excitations. The exciton character is also dominated by processes that rearrange spins on the Ni site, rather than moving charge between more extended states, which would tend to reduce the coupling of the exciton to phonons. These factors may contribute to the long lifetime and narrow linewidth of the exciton.

A key result of this study is that the \ce{NiPS3} exciton propagates like the double-magnon, even though the average energies of the exciton and double-magnon differ by more than one order of magnitude. This remarkable similarity can be understood by analyzing the exchange processes involved in the motion of the quasiparticles. We start by considering the spin-superexchange processes involved in exciton motion, finding that the exciton can swap its position with a spin (see Supplementary Note 4B for details). Consequently, when the exciton moves through the antiferromagnetic lattice, it generates a string of misaligned spins. Given that the exciton appears to propagate freely, we should consider processes that heal the misaligned spins in the wake of the exciton, which leads to the image in Fig.~\ref{fig:cartoon}a that illustrates the spin flips involved in exciton motion. Similar considerations can be applied to the motion of a double-magnon as shown in Fig.~\ref{fig:cartoon}b. Importantly, both exciton and double-magnon motion involve four spin exchanges. If we consider the sequence of different overlap integrals involved on the Ni and ligand states, the amplitude of the exciton hopping and double-magnon exchange processes are expected to be quite similar (see Supplementary Notes 4A and 4C). These considerations help us rationalize the similarities in the propagation of the two quasiparticles. A simple empirical tight-binding model fit to the exciton dispersion (see Supplementary Note 5) reveals that the third nearest neighbor interaction is the leading term in determining the exciton dispersion, consistent with the third nearest neighbor spin exchange being the dominant term in the spin Hamiltonian \cite{Scheie2023Spin, Wildes2022Magnetic}. We note in passing that a similar picture for a well-known free {\it single} magnon propagation in an antiferromagnet would require generating two spin exchanges \cite{Verresen2018quantum}.

The coupling between the exciton and magnetism might be relevant to why the $1.47$~eV exciton feature is visible in optics experiments. Since the exciton involves transitions between $d$-orbitals, it is expected to be nominally optically forbidden in a centrosymmetric crystal due to dipole selection rules. However, these rules can be lifted by perturbations that break the Ni-site symmetry, which includes exciton-spin or exciton-lattice interactions. The fact that there is a strong change in the optical cross section through $T_\text{N}$  \cite{Kim2018ChargeSpin}, whereas the \gls*{RIXS} is only modestly broadened, also supports the interpretation that exciton-spin interactions play a key role in the optical cross section.

Overall, our measurements reveal a Hund's excitonic quasiparticle in \ce{NiPS3} that propagates in a similar manner to a two-magnon excitation. Coming years will likely see further instrumental developments that allow \gls*{RIXS}, and exciton microscopy, measurement of \ce{NiPS3} to be extended to the ultrafast pump-probe regimes \cite{Dean2016ultrafast, Ginsberg2020spatially}. We believe that this has outstanding potential for understanding new means of using magnetic Hund's excitons to realize new forms of controllable transport of magnetic information.

\section{Methods \label{sec:methods}}
\subsection{Sample information \label{sec:methods_sample}}
\ce{NiPS3} bulk single crystal samples were procured from 2D Semiconductors, which synthesized the crystals by the chemical vapor transport method. The full unit cell of \ce{NiPS3} has a monoclinic symmetry (Space Group $C2/m$, \#12) with lattice parameters $a = 5.8$~\angstrom{}, $b = 10.1$~\angstrom{}, $c = 6.6$~\angstrom{}, and $\beta =  107.0^\circ$. We adopted this monoclinic-unit-cell convention, and index reciprocal space using scattering vector $\Q{}=(H,K,L)$ in reciprocal lattice units (r.l.u.). Therefore, the reciprocal lattice vector $\bm c^*$ is perpendicular to the $ab$ plane.

\ce{Ni^2+} ions in \ce{NiPS3} lie on a honeycomb lattice in the $ab$ plane and form with ABC-type stacking of the layers. Such stacking breaks the three-fold rotational symmetry of the monolayer structure which can be detected by measuring structural Bragg peaks such as ($0,2,4$). NiPS$_3$ is prone to characteristic twinning involving three equivalent domains rotated by $120^\circ$ in the $ab$ plane \cite{Wildes2015Magnetic,Wildes2022Magnetic}. Laboratory single-crystal x-ray diffraction measurements confirmed the presence of three twin domains in our samples. Therefore, measured quantities should be a weighted average over the three twin domains. We included these domain-averaging effects in the \gls*{ED} and magnon energy calculations. The apparent similarity in the measured dispersion between the two distinct directions (i.e., antiferromagnetic across the zig-zag chain and ferromagnetic along the chain) can be ascribed to the domain averaging effect due to the presence of structural twinning in the measured sample.

\ce{NiPS3} orders magnetically below a N\'{e}el temperature of $T_{\rm N}=159$~K \cite{Wildes2015Magnetic}. The magnetic unit cell is the same as the structural unit cell. It has a collinear magnetic structure consisting of antiferromagnetically coupled zigzag chains. 

\subsection{\gls*{RIXS} measurements \label{sec:methods_RIXS}}
Ultra-high-energy-resolution \gls*{RIXS} measurements were performed at the SIX 2-ID beamline of the National Synchrotron Light Source II \cite{Dvorak2016Towards}. The surface normal of the sample was $\bm c^*$ axis (i.e., $L$ direction). The in-plane orientation was determined by Laue diffraction. Then the pre-aligned sample was cleaved with Scotch tape in air to expose a fresh surface and immediately transferred into the \gls*{RIXS} sample chamber. Ni $L_3$-edge \gls*{RIXS} measurements were taken with linear horizontal ($\pi$) polarization with a scattering plane of either $(H0L)$ or $(0KL)$. The main resonance energy (around $853$~eV) is common for Ni-containing compounds \cite{Nakamura2016determining, Hepting2021soft, Wang2000nickel} but different from the previous report \cite{Kang2020Coherent}. This difference comes from the absolute energy calibration of the beamline, but does not affect the \gls*{RIXS} measurements and interpretations, which depend only on the relative changes. The spectrometer was operated with an ultrahigh energy resolution of $31$~meV \gls*{FWHM}. The temperature of the sample was kept at $T=40$~K except for the temperature-dependent measurements. Since the interlayer coupling in \ce{NiPS3} is relatively weak, the dispersion measurement was taken at a scattering angle of $2\Theta=150^\circ$ while varying the incident angle of the x-rays. A self-absorption correction \cite{Miao2017high} was applied to the \gls*{RIXS} spectra, which, however, does not affect peak positions. 

\subsection{Fitting of the \gls*{RIXS} spectra \label{sec:methods_fitting}}
In order to quantify the dispersion of the excitons and double-magnons, we fitted these peaks in the measured \gls*{RIXS} spectra to extract their peak positions. Although, in principle both the exciton and double-magnon can contain detailed substructure, we found that both features in our spectra can be accurately fit with simple peak shapes. 

In the double-magnon region, we used a Gaussian function for the elastic peak, a \gls*{DHO} model for both the magnon and double-magnon peaks, and a constant background. The width of the elastic peak was fixed to the energy resolution, which was determined by a reference measurement on a multilayer heterostructure sample designed to produce strong elastic scattering. The \gls*{DHO} equation for \gls*{RIXS} intensity $S(\Q{},\omega)$ as a function of $\Q{}$ and energy $\omega$ is
\begin{equation}
    S(\Q{},\omega)=\frac{\omega \chi_Q}{1-\exp(-\omega/k_B T)}\cdot \frac{2z_Q f_Q}{(\omega^2-f_Q^2)^2+(\omega z_Q)^2}
\end{equation}
where $f_Q$ is the undamped energy, $\chi_Q$ is the oscillator strength, $z_Q$ is the damping factor, $k_B$ is the Boltzmann constant, and $T$ is temperature. This \gls*{DHO} was then convoluted with a resolution function (a Gaussian function with peak width fixed to the energy resolution) to describe the magnons and double-magnons. We used the fitted value of $f_Q$ to represent the magnon or double-magnon peak energy and used $z_Q$ to characterize the peak width. 

In the exciton region, we resolved both the main exciton and an additional exciton sideband separated by $\sim 40$~meV as shown in Supplementary Fig.~3 (e.g., the spectrum at $K=-1.22$~r.l.u.), consistent with a previous report \cite{Kang2020Coherent}. Therefore, we fitted the data with two Voigt peaks with a third order polynomial background. The width of the Gaussian component was fixed to the energy resolution, and we constrained the widths of these two peaks to be the same. For the scans where the minor peak is not obvious, we further fixed the spacing between these peaks to the average value of $40$~meV obtained from other scans.

\subsection{Exact diagonalization RIXS calculations \label{sec:methods_ED}}

Our \ce{NiPS3} data were interpreted using standard \gls*{ED} methods for computing the  \gls*{RIXS} intensity \cite{Ament2011resonant}. The Kramers-Heisenberg formula for the cross section was used. This is derived by treating the interaction between the photon and the material within second order perturbation theory (as is required for scattering via an intermediate state resonance). We use the polarization-dependent dipole approximation for the photon absorption and emission interactions and simulate the presence of a core hole in the intermediate state with a core hole potential. In strongly correlated insulators like \ce{NiPS3}, accurate treatment of the electron-electron interactions are particularly important and the brief presence of a core hole means that local processes dominate the scattering. These factors means that cluster approximations are particularly appropriate and are widely used for this reason \cite{Ament2011resonant}. We therefore perform calculations for a \ce{NiS6} cluster, which can be projected onto \gls*{AIM} with essentially no loss of accuracy. As explained in detail in Supplementary Note 1, we were able to extract a well-constrained effective model for \ce{NiPS3} from the data. This was used to generate Fig.~\ref{fig:energy_map}b based on the parameters specified in Tab.~\ref{table:ED_parameters}. The detailed definitions of these parameters can be found in Supplementary Note 1. 

Since the \gls*{ED} method employed involves directly computing wavefunctions, the model can be used to extract the wavefunctions plotted in Fig.~\ref{fig:energy_map}g,h as outlined in Supplementary Note 2. All the calculations were done using the open-source software EDRIXS  \cite{Wang2019EDRIXS}.

\begin{table*}
\caption{\textbf{Full list of parameters used in the \gls*{AIM} calculations. }Units are eV.}
\begin{ruledtabular}
\begin{tabular}{cccccccccccccccccc}
$10Dq$ & $\epsilon_{p}$ & $V_{pd\sigma}$ & $V_{pp\sigma}$ & $F^0_{dd}$ & $F^2_{dd}$ & $F^4_{dd}$ & $F^0_{LL}$ & $F^2_{LL}$ & $F^4_{LL}$ & $U_{dL}$ & $F^0_{dp}$ & $F^2_{dp}$ & $G^1_{dp}$ & $G^3_{dp}$ & $\zeta_{\mathrm{i}}$ & $\zeta_{\mathrm{n}}$ & $\zeta_{\mathrm{c}}$ \\
\hline
0.42 & 7.5 & 0.95 & 0.8 & 7.88 & 10.68 & 6.68 & 0.46 & 2.59 & 1.62 & 1 & 7.45 & 6.56 & 4.92 & 2.80 & 0.083 & 0.102 & 11.4 \\
\end{tabular}
\end{ruledtabular}
\label{table:ED_parameters}
\end{table*}

\subsection{Magnetic cross-section calculations \label{sec:methods_CS}}

With the validated model in hand, it can be used to compute the resonance profile of the magnetic cross-section. A small Zeeman interaction was applied to the total spin angular momentum of the system, serving as the effective molecular magnetic field in the magnetically ordered state. The initially degenerate ground triplet consequently splits into three levels separated based on the spin state. After diagonalizing the Hamiltonian matrix, we used the Kramers-Heisenberg formula in the dipole approximation to calculate the incident-energy dependent \gls*{RIXS} cross section for transitions between the different elements of the triplet. The experimental geometry was explicitly included in the calculations. Three-fold twinning was also accounted for, which in fact has no effect on the final results due to the preserved cubic symmetry. We also note that these results are independent of the magnitude of the applied effective molecular magnetic field since this interaction is much smaller than the splitting between the ground state triplet and the lowest energy $dd$-excitation.

\section*{Data availability}
The \gls*{RIXS} data generated in this study have been deposited in the Zenodo database under access code 10791076 \cite{He2024data}. 

\section*{Code availability}
The calculations in this study were performed using the open source code EDRIXS \cite{Wang2019EDRIXS}. 

\section*{References}
\bibliographystyle{naturemag}
\bibliography{refs}
\clearpage

\section*{Acknowledgments}
We thank Sasha Chernyshev, Andy Millis, and Tatsuya Kaneko for fruitful discussions. Work performed at Brookhaven National Laboratory and Harvard Unversity was supported by the U.S.\ Department of Energy (DOE), Division of Materials Science, under Contract No.~DE-SC0012704. Theoretical calculations by K.W.\ was supported by the Excellence Initiative of the University of Warsaw (the `New Ideas' programme, decision No.~501-D111-20-2004310). E.B.\ was supported by the United States Army Research Office (W911NF-23-1-0394). S.~J.\ acknowledges support from the National Science Foundation under Grant No.~DMR-1842056. This research used beamline 2-ID of the National Synchrotron Light Source II, a U.S.\ DOE Office of Science User Facility operated for the DOE Office of Science by Brookhaven National Laboratory under Contract No.~DE-SC0012704.

\section*{Author contributions}
The project was conceived by M.P.M.D. W.H., Y.S., J.S., J.L., J.P., E.B., V.B., and M.P.M.D.\ performed the measurements. W.H., Y.S., K.W., M.W., S.J. E.B., M.M., and M.P.M.D.\ interpreted the data and performed the calculations. The paper was written by W.H., Y.S., and M.P.M.D. with input from all co-authors.

\section*{Additional Information}
Correspondence and requests for materials should be addressed to W.H.\ and M.P.M.D.

\section*{Competing interests}
The authors declare no competing interests.

\end{document}


\title{Supplementary Information for: Magnetically Propagating Hund's\\exciton in van der Waals antiferromagnet NiPS$_3$
}

\author{W.~He}
\email{whe1@bnl.gov}
\author{Y.~Shen}
\author{K.~Wohlfeld}
\author{J.~Sears}
\author{J.~Li}
\author{J.~Pelliciari}
\author{M. Walicki}
\author{S.~Johnston}
\author{E.~Baldini}
\author{V.~Bisogni}
\author{M.~Mitrano}
\author{M.~P.~M.~Dean}
\email{mdean@bnl.gov}

\renewcommand{\figurename}{Supplementary Figure}
\renewcommand{\tablename}{Supplementary Table}
\renewcommand{\thesection}{Supplementary Note \arabic{section}}

\def\mathbi#1{\ensuremath{\textbf{\em #1}}}
\def\Q{\ensuremath{\mathbi{Q}}}

\newcommand\tlc[1]{\texorpdfstring{\lowercase{#1}}{#1}}

\maketitle

\tableofcontents

\section{Exact diagonalization calculations}

Here we outline the detailed approach and values used in our \gls*{RIXS} calculations. 
Throughout this section, we represent the model in hole (rather than electron) language. 

\subsection{Cluster model}
Our cluster representation of \ce{NiPS3} uses six S atoms surrounding a central Ni atom. The orbital basis includes the full set of ten Ni $3d$ spin-orbitals and six S $3p$ spin-orbitals for each of the six neighboring S ions. The Ni $2p_{3/2}$ core states are included to simulate the resonant pathway involved in \gls*{RIXS}. The Hamiltonian for this model is
\begin{equation}
\mathcal{H}=\hat{E}_{d}+\hat{E}_{p}+\hat{U}_{dd}+\hat{U}_{pp}+\hat{U}_{dp}+\hat{U}_{q}+\hat{T}_{pp}+\hat{T}_{dp}+\hat{\zeta},
\end{equation}
which includes the on-site energies $\hat{E}_{d}$ and $\hat{E}_{p}$ for the Ni $3d$ and S $3p$ orbitals, respectively. Here, $\hat{U}_{dd}$ and $\hat{U}_{pp}$ are on-site Coulomb interactions and $\hat{U}_{dp}$ are the intersite interactions, and $\hat{U}_{q}$ is the interaction between the core and valence holes. Regarding the intersite hybridization, $\hat{T}_{pp}$ indicates $p$--$p$ hopping among S sites and $\hat{T}_{dp}$ stands for $p$--$d$ hopping between Ni and S sites. The charge-transfer energy $\Delta$ is included implicitly as discussed below. $\hat{\zeta}$ represents spin-orbit coupling, which we include for the core and valence states. 

The crystal field associated with the trigonal distortion in \ce{NiPS3} is on the order of only 1~meV \cite{Afanasiev2021Controlling, Chandrasekharan1994Magnetism}, We therefore assumed cubic symmetry for simplicity. The $t_{\mathrm{2g}}-e_{\mathrm{g}}$-orbital energy splitting is further set by the Ni crystal field, and is specified by $10D_q$. To describe Ni-S hopping, we use the Slater-Koster parameters of $V_{pp\sigma}$, $V_{pp\pi}$, $V_{pd\pi}$, and $V_{pd\sigma}$, which denote hopping between Ni $d$ and S $p$ states with either $\pi$ or $\sigma$ orbital symmetry. We denote the energy of the S $3p$ orbitals measured relative to the Ni $e_g$ orbitals by $\epsilon_p$ and neglect the splitting between the $3p_\sigma$ and $3p_\pi$ orbitals in the cluster. 

The Coulomb interactions in the model are specified using Slater integrals. Following standard methods, we include $F^0_{dd}$, $F^2_{dd}$, and $F^4_{dd}$ to describe interactions on Ni $3d$ orbitals. Similarly, we used $F^0_{pp}$ and $F^2_{pp}$ for S $3p$ orbitals. $F^0_{dp}$ is the intersite Coulomb interaction between Ni $3d$ and S $3p$ orbitals. In the \gls*{RIXS} intermediate states, we used $F^0_{dp}$, $F^2_{dp}$, $G^1_{dp}$, and $G^3_{dp}$ to describe the Coulomb interactions between Ni $3d$ and Ni $2p$ holes. Lastly, we also included spin-orbit coupling terms of Ni $3d$ orbitals for the initial and intermediate states ($\zeta_{\mathrm{i}}$ and $\zeta_{\mathrm{n}}$), as well as the much larger core-hole coupling ($\zeta_{\mathrm{c}}$).  An inverse core-hole lifetime $\Gamma_c = 0.6$~eV \gls*{HWHM} was used in order to fit the observed width of the resonance and the final state energy loss spectra are broadened using a Gaussian function with a \gls*{FWHM} of $0.05$~eV, in order to match the observed width of the final states.

\subsection{Determining the cluster model parameters}

\begin{table*}
\caption{\textbf{Full list of parameters used for the NiS$_6$ cluster model in the ED calculations.} Units are eV.}
\begin{ruledtabular}
\begin{tabular}{cccccccccccc}
\multicolumn{6}{c}{On-site orbital energies and spin-orbit coupling} & & \multicolumn{5}{c}{On-site Coulomb interactions} \\
$\epsilon_{d}(e_{\mathrm{g}})$ & $\epsilon_{d}(t_{\mathrm{2g}})$ & $\epsilon_p$ & $\zeta_{\mathrm{i}}$ & $\zeta_{\mathrm{n}}$ & $\zeta_{\mathrm{c}}$ & & $F^0_{dd}$ & $F^2_{dd}$ & $F^4_{dd}$ & $F^0_{pp}$ & $F^2_{pp}$ \\
0 & 0.42 & 7.5 & 0.083 & 0.102 & 11.4 & & 7.88 & 10.68 & 6.68 & 3.2 & 5 \\
\hline
\multicolumn{2}{c}{Hopping integrals} & & \multicolumn{4}{c}{Intersite Coulomb interactions} & & \multicolumn{4}{c}{Core-hole potential}\\
$V_{pd\sigma}$ & $V_{pp\sigma}$ & & $F^0_{dp}$ & $F^2_{dp}$ & $G^1_{dp}$ & $G^3_{dp}$ & & $F^0_{dp}$ & $F^2_{dp}$ & $G^1_{dp}$ & $G^3_{dp}$ \\
0.95 & 0.8 & & 1 & 0 & 0 & 0 & & 7.45 & 6.56 & 4.92 & 2.80 \\
\end{tabular}
\end{ruledtabular}
\label{table:ED_NiS6}
\end{table*}

\begin{figure}
\includegraphics{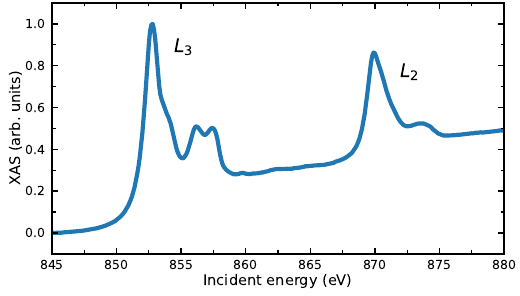}
\caption{\textbf{\gls*{XAS} spectra}. The spectra was taken at 40~K using $\pi$-polarized x-rays in the total fluorescence yield mode. The sample geometry was the same as the measurement for the \gls*{RIXS} energy map (i.e., $\theta = 22.6^\circ$, $2\Theta = 150^\circ$, $(0KL)$ scattering plane) except that the photo diode detector was placed slightly above the scattering plane. }
\label{fig:SI_XAS}
\end{figure}

Although there are several parameters in the original Hamiltonian, they can be constrained by physical considerations and by exploiting the richness of the \ce{NiPS3} \gls*{RIXS} spectra. We start by outlining the different constraints on the parameters:

\begin{enumerate}

\item \textbf{Coulomb interactions}: The Slater integrals for Ni $3d$ Coulomb interactions can be recast into onsite Coulomb repulsion $U$ and Hund's coupling $J_H$. Since we work in the hole language, the on-site Coulomb repulsion for the ligands is not crucial since double hole occupation on ligands is unlikely. Indeed, we get nearly zero probability for the $d^{10}\underline{L}^2$ ($\underline{L}$ stands for a ligand hole) configuration in our calculations, as shown in Fig.~1e in the main text. Therefore, $F^0_{pp}$ and $F^2_{pp}$ do not influence our conclusions and are fixed to standard values appearing in the literature~\cite{Wang2019EDRIXS}. Due to the distance between the Ni and S atoms the strength of the intersite Coulomb interaction $F^0_{dp}$ is expected to be much smaller than the Ni on-site Coulomb interactions and thus plays a negligible role. We therefore fix this value to 1~eV, consistent with values often found in the literature~\cite{Neudert2000fourband}. The precise value of this interaction is not crucial because its effect on the charge-transfer energy can also be absorbed into the S $3p$ orbital energy $\epsilon_p$. 

\item \textbf{Hopping integrals}: Making use of the Slater-Koster scheme, all the hopping integrals can be derived from two parameters, $V_{pd\sigma}$ and $V_{pp\sigma}$. Here, we fixed $V_{pd\pi}=-V_{pd\sigma}/2$ and $V_{pp\pi}=-V_{pp\sigma}/4$, which is common for transition metal compounds~\cite{Mizuno1998electronic}.

\item \textbf{Charge-transfer energy}: In a multi-orbital system, this energy is defined as $\Delta=\epsilon(d^9\underline{L})-\epsilon(d^8)=[\epsilon_d(e_{\mathrm{g}})+\epsilon_p+F^0_{dp}]-[\epsilon_d(e_{\mathrm{g}})\times 2+F^0_{dd}-8F^2_{dd}/49-F^4_{dd}/49]$. Once the Ni $3d$ on-site Coulomb interactions and $10D_q$ are chosen, the charge-transfer energy is only affected by the energy difference between the S $2p$ states and the Ni $3d$ $e_g$ states, i.e. $\epsilon_p$ (recall we set the latter to zero).

\item \textbf{Core-hole potential}: The core-hole interactions are known to be only weakly screened in the solid state. We initially set the values to 80\% of their atomic values and further refined them based on the \gls*{XAS} (Supplementary Fig.~\ref{fig:SI_XAS}) after we determined other above parameters from the \gls*{RIXS} spectra.

\item \textbf{Spin-orbit coupling}: The spin-orbit coupling terms for the Ni $3d$ orbitals are weak and have negligible effects of the spectra. We consequently simply fixed them to their atomic values. The core-hole spin-orbit coupling parameter is adjusted slightly to match the $L_2$ peak position. 
\end{enumerate}

Aside from the core-hole potential, we have $10D_q$, $U$, $J_H$, $V_{pd\sigma}$, $V_{pp\sigma}$, and $\epsilon_p$ as the only free parameters, which have distinct effects on the \gls*{RIXS} spectra. The energy of the excitation around $1.0$~eV is mostly sensitive to $J_H$, while the position of the feature around $1.1$~eV is primarily determined by $10D_q$. The exciton energy is correlated with a combination of $J_H$, $10D_q$, and $U$. Once $J_H$ and $10D_q$ are determined by the excitations around $1.0$ and $1.1$~eV, respectively, we can tune the onsite Coulomb repulsion $U$ to match the exciton energy. $\epsilon_p$ is constrained by the resonant energy as well as the intensity of the exciton, while the hopping integrals can be determined mainly by the resonant energy of satellite peaks. Using the above strategy, we successfully determine these parameters with estimated error bars of $\sim1$~eV for all the Coulomb interactions and $\sim0.2$~eV for hopping integrals and $10D_q$. The final parameters are provided in Tab.~\ref{table:ED_NiS6}.

\subsection{Anderson impurity model}

The NiS$_6$ cluster model can be effectively projected to an \gls*{AIM} using ligand field theory~\cite{Eskes1990cluster,Haverkort2012multiplet}. The S $3p$ orbitals form 10 ligand orbitals that have the same symmetry as the Ni $3d$ orbitals through $p$--$p$ hopping. The ligand orbitals further hybridize with the Ni $3d$ orbitals, which can be parameterized based on $p$--$d$ hopping integrals. Correspondingly, the energies of the ligand orbitals and their hopping integrals with the Ni $3d$ orbitals can be evaluated as shown in Tab.~\ref{table:ligand}. The missing piece is the onsite Coulomb interactions for the ligand orbitals, and we refined these interactions using the same approach as for the cluster model. The full list of the parameters are shown in Tab.~\ref{table:ED_AIM}. We will show later that the \gls*{AIM} calculation results are, as expected, essentially identical to the NiS$_6$ cluster results.

\begin{table}
\caption{\textbf{Projection of the NiS$_6$ cluster model to the \gls*{AIM}.} Here, $\epsilon_L$ refers to the ligand orbital energies determined by the S $3p$ orbital energy $\epsilon_p$ and $T_{pp}=|V_{pp\sigma}-V_{pp\pi}|$. The hopping integrals $V$ between Ni $3d$ orbitals and ligand orbitals can be evaluated from $p$--$d$ hoppings. Units are eV.}
\begin{ruledtabular}
\begin{tabular}{cccc}
Symmetry & $\epsilon_d$ & $\epsilon_L$ & $V$ \\
\hline
$e_{\mathrm{g}}$  & $\epsilon_{d}(e_{\mathrm{g}})$ & $\epsilon_{p}-T_{pp}$ & $\sqrt{3}|V_{pd\sigma}|$ \\
$t_{\mathrm{2g}}$  & $\epsilon_{d}(t_{\mathrm{2g}})$ & $\epsilon_{p}+T_{pp}$ & $2|V_{pd\pi}|$ \\
\end{tabular}
\end{ruledtabular}
\label{table:ligand}
\end{table}

\begin{table}
\caption{\textbf{Full list of parameters used for the AIM in the ED calculations.} Here, $\zeta_{\mathrm{L}}$ is the spin-orbit coupling strength for the ligand orbitals, $U_{dL}$ is intersite Coulomb interactions, and $F^0_{LL}$, $F^2_{LL}$, and $F^4_{LL}$ are the Slater integrals for ligand orbital onsite Coulomb interactions. Units are eV.}
\begin{ruledtabular}
\begin{tabular}{cccccccccccc}
\multicolumn{4}{c}{On-site orbital energies} & & \multicolumn{4}{c}{Spin-orbit coupling} & & & Hopping integral \\
$\epsilon_{d}(e_{\mathrm{g}})$ & $\epsilon_{d}(t_{\mathrm{2g}})$ & $\epsilon_{L}(e_{\mathrm{g}})$ & $\epsilon_{L}(t_{\mathrm{2g}})$ & & $\zeta_{\mathrm{i}}$ & $\zeta_{\mathrm{n}}$ & $\zeta_{\mathrm{c}}$ & $\zeta_{\mathrm{L}}$ & & & $V_{pd\sigma}$ \\
0 & 0.42 & 6.5 & 8.5 & & 0.083 & 0.102 & 11.4 & 0 & & & 0.95 \\
\hline
\multicolumn{7}{c}{Coulomb interactions} & & \multicolumn{4}{c}{Core-hole potential} \\
$F^0_{dd}$ & $F^2_{dd}$ & $F^4_{dd}$ & $F^0_{LL}$ & $F^2_{LL}$ & $F^4_{LL}$ & $U_{dL}$ & & $F^0_{dp}$ & $F^2_{dp}$ & $G^1_{dp}$ & $G^3_{dp}$ \\
7.88 & 10.68 & 6.68 & 0.46 & 2.59 & 1.62 & 1 & & 7.45 & 6.56 & 4.92 & 2.80 \\
\end{tabular}
\end{ruledtabular}
\label{table:ED_AIM}
\end{table}

\subsection{Single site atomic model}

To test whether a simpler model might be able to capture the relevant physics, we also considered an atomic model composed of a single Ni atom with 10 effective Ni $3d$ orbitals. In this case, these spin orbitals represent effective hybridized Ni-S orbitals, so they are consequently quite different from the Ni orbitals in the above models. As expected, with appropriate values of crystal field and Hund's coupling, this model has a triplet ground state and a singlet excited state. We refined the parameters using a methodology similar to the one we used for our cluster model. However, we found that the exciton energy could not be accurately reproduced with this model, which we will discuss in more depth when we compare the models. The obtained values of these parameters are listed in Tab.~\ref{table:ED_atomic}.

\begin{table}
\caption{\textbf{Full list of parameters used for the single site atomic model in the ED calculations.} All parameters are in units of eV.}
\centering
\begin{ruledtabular}
\begin{tabular}{cccccccccccc}
$\epsilon_{d}(e_{\mathrm{g}})$ & $\epsilon_{d}(t_{\mathrm{2g}})$ & $\zeta_{\mathrm{i}}$ & $\zeta_{\mathrm{n}}$ & $\zeta_{\mathrm{c}}$ & $F^0_{dd}$ & $F^2_{dd}$ & $F^4_{dd}$ & $F^0_{dp}$ & $F^2_{dp}$ & $G^1_{dp}$ & $G^3_{dp}$\\ 
\hline
0 & 1.07 & 0.083 & 0.102 & 11.2 & -0.70 & 5.26 & 3.29 & 0.26 & 6.18 & 2.89 & 1.65 \\
\end{tabular}
\end{ruledtabular}
\label{table:ED_atomic}
\end{table}

\subsection{Discussion of an alternative model}

Reference \cite{Kang2020Coherent} (which we will refer to as Ref. ``A'') used a NiS$_6$ cluster to calculate the \gls*{RIXS} spectra as we did here but adopting different parameters. Most of the parameters are listed in that paper but two of them are not, namely the energy different between Ni $e_{\mathrm{g}}$ and S $3p$ orbitals and the core-hole potential parameter $F^0_{dp}$, which can be indirectly inferred from $\Delta=\epsilon(d^9\underline{L})-\epsilon(d^8)=0.95$~eV and $\Delta'=\epsilon(d^{10}\underline{c}\underline{L})-\epsilon(d^9\underline{c})=-0.55$~eV where $\underline{c}$ stands for a core hole. It can be easily concluded that $\epsilon_p-\epsilon_d(e_{\mathrm{g}})=6.37$~eV but it is tricky to obtain $F^0_{dp}$ since it depends on how core-hole potential is included in $\Delta'$. We choose it to be $7.79$~eV. Similarly, we inferred that a similar (but not identical) inverse core-hole lifetime of $0.4$~eV \gls*{HWHM} was used and their spectra were broadened using a Lorentzian function with \gls*{FWHM} of $0.1$~eV. Table~\ref{table:ED_NiS6_Nature} lists all the parameters. We have confirmed that our approach reproduces Ref.~A's results \cite{Kang2020Coherent} when we adopt their parameters. 

\begin{table}
\caption{\textbf{Full list of parameters used for the NiS$_6$ cluster model in Ref.~A \cite{Kang2020Coherent}.} The italic values are parameters that are not directly listed in the paper but inferred from other parameters. Units are eV.}
\begin{ruledtabular}
\begin{tabular}{cccccccccccc}
\multicolumn{6}{c}{On-site orbital energies and spin-orbit coupling} & & \multicolumn{5}{c}{On-site Coulomb interactions} \\
$\epsilon_{d}(e_{\mathrm{g}})$ & $\epsilon_{d}(t_{\mathrm{2g}})$ & $\epsilon_p$ & $\zeta_{\mathrm{i}}$ & $\zeta_{\mathrm{n}}$ & $\zeta_{\mathrm{c}}$ & & $F^0_{dd}$ & $F^2_{dd}$ & $F^4_{dd}$ & $F^0_{pp}$ & $F^2_{pp}$ \\
0 & 1 & $\mathit{6.37}$ & 0.08 & 0.08 & 11.5 & & 8 & 14.68 & 9.12 & 1 & 0 \\
\hline
\multicolumn{2}{c}{Hopping integrals} & & \multicolumn{4}{c}{Intersite Coulomb interactions} & & \multicolumn{4}{c}{Core-hole potential}\\
$V_{pd\sigma}$ & $V_{pd\pi}$ & & $F^0_{dp}$ & $F^2_{dp}$ & $G^1_{dp}$ & $G^3_{dp}$ & & $F^0_{dp}$ & $F^2_{dp}$ & $G^1_{dp}$ & $G^3_{dp}$ \\
0.74 & -0.4 & & 0 & 0 & 0 & 0 & & $\mathit{7.79}$ & 6.18 & 4.63 & 2.63 \\
\end{tabular}
\end{ruledtabular}
\label{table:ED_NiS6_Nature}
\end{table}

\subsection{Comparison of all the models}

\begin{figure*}
\includegraphics{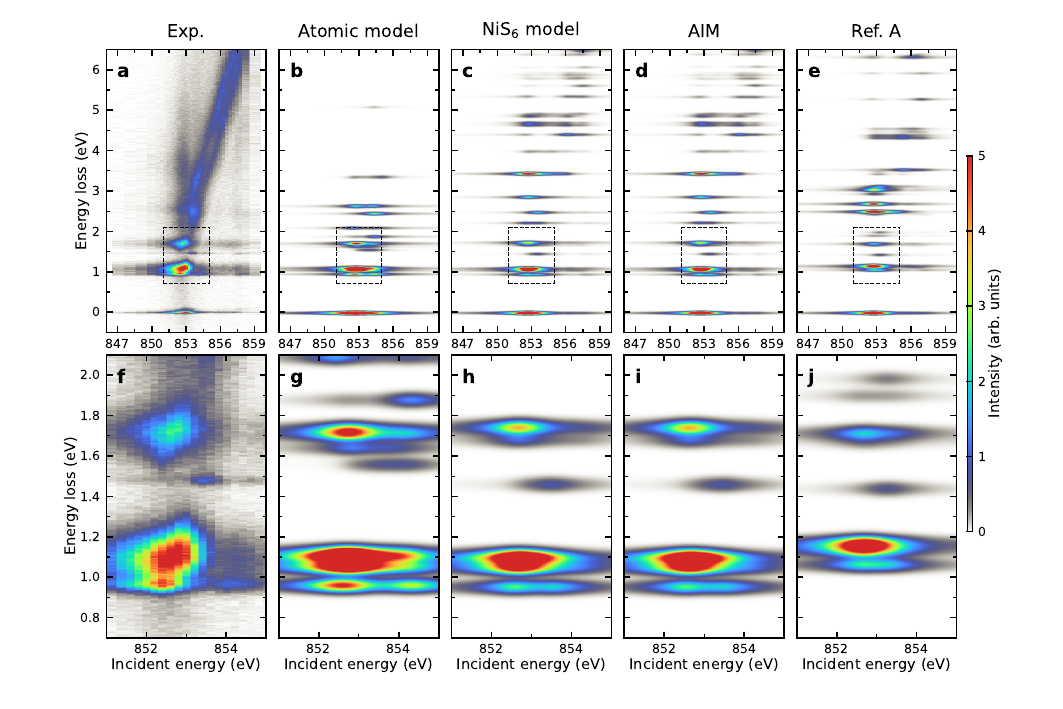}
\caption{\textbf{Comparison of \gls*{RIXS} energy maps calculated using different models}. \textbf{a--e}, The measured and calculated \gls*{RIXS} energy maps. \textbf{f--j}, Zoom-in of the measured and calculated \gls*{RIXS} maps in the region outlined by the dashed box in panels \textbf{a--e}. The calculations assume a constant lifetime for the excitations, and omit the continuum of states needed to accurately simulate the fluorescent processes that occur at high energies, so the transitions above 2~eV appear far narrower in the calculations compared to the theory. }
\label{fig:SI_ED_Emaps}
\end{figure*}

Supplementary Fig.~\ref{fig:SI_ED_Emaps} summarizes the calculation results for all the three models, as well as the one from Ref.~A \cite{Kang2020Coherent}. The exciton at around $1.47$~eV can only be captured in the models with ligand orbitals. Compared with the combination of parameters used in Ref.~A \cite{Kang2020Coherent}, our model (NiS$_6$ cluster or equivalently \gls*{AIM}) better captures the $dd$ excitations at 1.0 and 1.1~eV. We also see that our model has the secondary satellite resonance at 857.7~eV, consistent with the data, whereas the model of Ref.~A \cite{Kang2020Coherent} predicts this feature around 857~eV. Moreover, our model is compatible with the previous optical work, which reported absorption peaks at $1.1$, $1.7$, $2.2$, $3.5$, and $4.6$~eV \cite{Kim2018ChargeSpin}. On the contrary, the calculated spectra using the model in Ref.~A \cite{Kang2020Coherent} cannot reproduce the energies of the last three excitations well.

If one considers the exciton in isolation, without requiring a precise reproduction of the shape of the $dd$-excitations, our model and Ref.~A \cite{Kang2020Coherent} both obtain 1.47 eV consistent with the experimental energy of this mode. This observation suggests that there is some ambiguity in determining the model parameters from the exciton energy alone. By analyzing the different models, we found that this ambiguity arises from an anticorrelation between Hund's interactions and charge transfer energy, which produces a ranges of parameters that obtain the observed exciton energy. We note, however, that the Hund's energy is known to be weakly screened in solids and generally falls in a reasonably narrow range of values around 70-90\% of the values calculated for isolated atoms \cite{Khomskii2014transition}. In fact, the onsite Coulomb interaction parameters for Ni $3d$ orbitals in Ref.~A \cite{Kang2020Coherent}, $F^2_{dd}$ and $F^4_{dd}$, are 120\% of atomic values, which is unphysical. Our parameters are 87\% of atomic values, which falls within the normal range for correlated insulators. 

Although the ground state and exciton in all these models share similar overall symmetry, their electronic composition varies appreciably. For example, our charge-transfer energy ($2.50$~eV) is larger than that reported in Ref.~A ($0.95$~eV) \cite{Kang2020Coherent}. Correspondingly, the exciton in our model, which is a singlet, has $1.26$ versus $0.74$ holes occupying the Ni and S sites, respectively. Conversely, it has $1.08$ versus $0.92$ holes occupying the Ni and S sites, respectively, in the model of Ref.~A \cite{Kang2020Coherent}. As noted in the main text, the preferential Ni occupation of the exciton is vital for understanding the interactions underlying the exciton formation. 

\section{Wavefuction analysis}

As discussed in the main text, we performed a detailed wavefunction analysis for the \gls*{AIM} model to investigate the nature of the exciton, but results are similar for both the \gls*{AIM} and cluster models.

Figure~1\textbf{d}--\textbf{f} plots several expectation values for the \ce{NiPS3} wavefunction, which reveals a significant portion of doubly occupied $d^8$ configuration and deviation from the pure $d^9\underline{L}$ states. A full understanding of these effects can be obtained by projecting the real wavefunctions onto those devised by Zhang and Rice to describe the $d^9\underline{L}$ manifold \cite{Zhang1988Effective}. The Zhang-Rice singlet can be written as
\begin{align}
|\mathrm{ZRS}\rangle & = \frac{1}{2}(|d_{x^2-y^2, \downarrow}L_{x^2-y^2, \uparrow}| - |d_{x^2-y^2, \uparrow}L_{x^2-y^2, \downarrow}| + |d_{z^2, \downarrow}L_{z^2, \uparrow}| - |d_{z^2, \uparrow}L_{z^2, \downarrow}|). 
\end{align}
The Zhang-Rice triplet states are
\begin{equation}
\begin{split}
|\mathrm{ZRT}, S_z=1\rangle &= \frac{1}{\sqrt{2}}(|d_{x^2-y^2, \uparrow}L_{z^2, \uparrow}| - |d_{z^2, \uparrow}L_{x^2-y^2, \uparrow}|) \\
|\mathrm{ZRT}, S_z=0\rangle &= \frac{1}{2}(|d_{x^2-y^2, \uparrow}L_{z^2, \downarrow}| + |d_{x^2-y^2, \downarrow}L_{z^2, \uparrow}| - |d_{z^2, \uparrow}L_{x^2-y^2, \downarrow}| - |d_{z^2, \downarrow}L_{x^2-y^2, \uparrow}|) \\
|\mathrm{ZRT}, S_z=-1\rangle &= \frac{1}{\sqrt{2}}(|d_{x^2-y^2, \downarrow}L_{z^2, \downarrow}| - |d_{z^2, \downarrow}L_{x^2-y^2, \downarrow}|).
\end{split}
\end{equation}
Since $d^8$ states have a significant amount of weight in the wavefunction, these states must also be considered. We therefore define 
\begin{equation}
\begin{split}
|^1A_1, d\rangle &= \frac{1}{\sqrt{2}}(|d_{x^2-y^2, \uparrow}d_{x^2-y^2, \downarrow}| + |d_{z^2, \uparrow}d_{z^2, \downarrow}|) \\
|^3A_2, d, S_z=1\rangle &= |d_{x^2-y^2, \uparrow}d_{z^2, \uparrow}| \\
|^3A_2, d, S_z=0\rangle &= \frac{1}{\sqrt{2}}(|d_{x^2-y^2, \downarrow}d_{z^2, \uparrow}| + |d_{x^2-y^2, \uparrow}d_{z^2, \downarrow}|) \\
|^3A_2, d, S_z=-1\rangle &= |d_{x^2-y^2, \downarrow}d_{z^2, \downarrow}| .
\end{split}
\end{equation}
Similar states exist with two ligand holes as
\begin{equation}
\begin{split}
|^1A_1, L\rangle &= \frac{1}{\sqrt{2}}(|L_{x^2-y^2, \uparrow}L_{x^2-y^2, \downarrow}| + |L_{z^2, \uparrow}L_{z^2, \downarrow}|) \\
|^3A_2, L, S_z=1\rangle &= |L_{x^2-y^2, \uparrow}L_{z^2, \uparrow}| \\
|^3A_2, L, S_z=0\rangle &= \frac{1}{\sqrt{2}}(|L_{x^2-y^2, \downarrow}L_{z^2, \uparrow}| + |L_{x^2-y^2, \uparrow}L_{z^2, \downarrow}|) \\
|^3A_2, L, S_z=-1\rangle &= |L_{x^2-y^2, \downarrow}L_{z^2, \downarrow}| .
\end{split}
\end{equation}
Using this basis, the ground state triplet can be described in terms of the $d^8$ triplet state mixed with a Zhang-Rice triplet, with only a small contribution from other states (which we denote $|\dots\rangle$). The wavefunctions are 
\begin{equation}
\begin{split}
    |\mathrm{GS}, S_z=-1\rangle &= 0.757\ |^3A_2, d, S_z=-1\rangle + 0.163\ |^3A_2, L, S_z=-1\rangle + 0.625\ |\mathrm{ZRT}, S_z=-1\rangle + 0.099\ |\dots\rangle \\
    |\mathrm{GS}, S_z=0\rangle &= 0.757\ |^3A_2, d, S_z=0\rangle + 0.163\ |^3A_2, L, S_z=0\rangle +0.625\ |\mathrm{ZRT}, S_z=0\rangle + 0.099\ |\dots\rangle \\
    |\mathrm{GS}, S_z=1\rangle &= 0.757\ |^3A_2, d, S_z=1\rangle + 0.163\ |^3A_2, L, S_z=1\rangle + 0.625\ |\mathrm{ZRT}, S_z=1\rangle + 0.099\ |\dots\rangle .
\end{split}
\end{equation}
The wavefunction of the exciton can be expressed as
\begin{equation}
    |\mathrm{E}\rangle = 0.430\ |^1A_1, d\rangle + 0.209\ |^1A_1, L\rangle + 0.766\ |\mathrm{ZRS}\rangle + 0.430\ |\dots\rangle, 
\end{equation}
which demonstrates that the exciton is primarily composed of the singlet components of the $d^8$, $d^9\underline{L}$, and $d^{10}\underline{L}^2$ states. We also see a noticeable increase in the ligand character compared to the ground state. A further component arises from mixing with a large number of other states. 

This wavefunction analysis can further be used to understand the interactions underlying the exciton energy. While the Zhang-Rice singlet and triplet are only split by quite weak exchange interactions, the exciton state has an appreciable fraction of doubly occupied holes, which involve much stronger Hund's exchange. To more accurately quantify the contribution from Hund's interaction $J_H$, we compute the expectation value of the operator $\hat{J}_{H}$ for the exciton wavefunction, i.e., $\langle E |\hat{J}_{H}| E \rangle$. The calculated expectation value is $1.60$~eV, which is indeed the leading factor to set the exciton energy scale and compensated slightly by the charge-transfer process to give the exciton energy of $1.47$~eV. 

\section{Detailed fitting procedures for the \gls*{RIXS} spectra}
\label{sec:RIXS_fits}

As noted in Methods Section, we use Voigt functions to fit the excitons, and use a \gls*{DHO} convoluted with a Gaussian resolution function to fit the magnon as well as the double-magnon peaks. In this section, we provide further details on our fitting procedures. 

\subsection{Energy zero determination}
To ensure a reliable measurement of the subtle exciton dispersion we observe here, we require an accurate calibration of the energy zero, which is often done by fitting the elastic line. However, in the current case of \ce{NiPS3} we only observe quite weak elastic scattering and it is only possible to precisely fit the elastic line in a subset of the spectra. We consequently use the magnon peak position to perform the fine alignment of the spectra. The expected magnon peak energy is calculated using spin wave theory \cite{Toth2015Linear} based on the weighted sum of the magnon branches predicted in the Hamiltonian obtained in Ref.~\cite{Scheie2023Spin}. In this process, structural domain averaging introduces an error of $\sim 1$~meV and the errors of the neutron results themselves are $\sim 2$~meV \cite{Scheie2023Spin, Wildes2022Magnetic}. For the spectra at $|H(K)| \leq 0.1$, because the magnon and elastic peaks are too close to each other, to make the fit converge, we have to manually fix the energy zero to the value that gives the lowest $\chi^2$. In this case, we assign an estimated error bar of $5$~meV for the energy zero because the fits would severely deviate from the measured lineshapes if we shifted the energy zero by $\pm5$~meV or more. The error bars for the fitted double-magnon and exciton peak positions shown in Fig.~2 include not only their own fitting error ($0.5$--$1$~meV for the exciton and $\sim 3$~meV for the double-magnon) but also the fitting error of the energy zero ($\sim 1$~meV) as well as the uncertainty of the calculated magnon energies (model differences and the twinning effect as explained above).

\subsection{Best fits to the low temperature momentum-dependent \gls*{RIXS} spectra}

We present the best fits to the linecuts of the \gls*{RIXS} spectra measured at various in-plane momentum transfer at $T=40$~K in Supplementary Figs.~\ref{fig:SI_LT_exciton}--\ref{fig:SI_0KL_LT_magnon}. Particularly for the low-energy region, we display the three components used in the fits to clarify the identify of different spectral features. 

\subsection{Cross-checks of the exciton and double-magnon dispersions}
To verify the fitted exciton dispersion, we performed three tests. The first one is to check the consistency with the reported zone-center exciton energy from optical measurements. The exciton energy from previous photoluminescence and optical absorption spectra studies is $1.475$~eV with uncertainty below 1~meV \cite{Kang2020Coherent}, which is in line with our fitting results. Even after taking into account the uncertainty from the energy calibration of our spectrometer ($\sim 2.5$~meV at this energy), this zone-center exciton energy taken from optical measurements still has smaller error bar than our fitting results in this region and can help to see an energy difference between the zone-center exciton and excitons at higher $\Q{}$ in Fig.~2. Since the fitted exciton energy heavily depends on the fitting quality of the low-energy region, we next inspect the fitted elastic and magnon peak intensities. As shown in Supplementary Fig.~\ref{fig:SI_fits_comparison}, the fitted magnon intensity matches calculated values quite well. The elastic peak becomes stronger near the zone center as expected for specular reflection. The last check we performed is to test the null hypothesis in which we presume that the exciton is, in fact, non-dispersive, and that the apparent dispersion comes from calibration errors. We find that this indeed leads to an unphysical result in which the elastic line intensity drops at the specular $(0, 0)$ position, when it would be expected to increase or stay the same (Supplementary Fig.~\ref{fig:SI_fits_comparison}).

Similarly, we also did the null hypothesis test for the double-magnons. Even though the double-magnon peak is broader than the exciton peak and therefore has larger fitting error for the peak positions, we can still exclude the null hypothesis after carefully inspecting the fitted curves. This is most evident in the spectra near the Brillouin zone center, where the fits assuming non-dispersive double-magnons clearly deviate from the best fits and fail to describe the experimental data (Supplementary Fig.~\ref{fig:SI_fits_comparison2}).

\subsection{Magnon and double-magnon fits with Voigt functions}
Although the \gls*{DHO} model are generally used to fit low-energy magnetic excitations in \gls*{RIXS} spectra, to avoid uncertainties inherent in the model we used, we also investigate the Voigt model to fit the magnons and double-magnons. The fitting result turns out to be quite similar to the original \gls*{DHO} fits (Supplementary Fig.~\ref{fig:SI_DHO_vs_Voigt}). Since the magnon peak is quite narrow, both \gls*{DHO} and Voigt models give nearly identical fitting results. Therefore, the change to the energy zero calibration is minimal, so as to the fitted exciton peak positions. For the double-magnon peak, Voigt function gives similar fitting quality but consistently lower peak positions (by $\sim2$~meV) as a consequence of increased peak width. However, the existence of the double-magnon dispersion and its resemblance to the exciton dispersion are still valid in the Voigt model fits. 

Despite that the low-temperature data set is unable to distinguish the two models, the \gls*{DHO} model is more appropriate to fit the high-temperature data set, where the anti-stoke peaks on the negative energy loss side become more apparent (Supplementary Fig.~\ref{fig:SI_H0L_HT_magnon}). For consistency, we therefore adopt the \gls*{DHO} model throughout the manuscript to fit the magnon and double-magnon peaks.

\subsection{Fits to the high-temperature spectra}
Fitting the high-temperature spectra at $190$~K is more difficult since the magnon peak is softened and cannot be used for energy zero determination. Thanks to the enhanced elastic peak, we are able to fit the spectra at several large $\Q{}$ positions (e.g., $H=0.69$~r.l.u.) where the elastic, magnon, and double-magnon peaks are well separated. The magnon peak is not only softened, but also broadened compared to the low temperature data. We then fix its damping factor $z_Q$ to an average value of $16$~meV. Supplementary Fig.~\ref{fig:SI_H0L_HT_magnon} is the best fits we obtained for the low-energy region, although the error bars for the energy zero determination could be as large as $\sim 5$~meV. Then we fit the exciton peak as shown in Supplementary Fig.~\ref{fig:SI_HT_exciton}, which has a clear softening and broadening. However, the existence of a dispersive mode is ambiguous here due to the large error bars on the fitted peak positions.

\subsection{Fits to the incident energy dependent spectra}
To quantify the resonant behaviors of excitons and double-magnons, we also fit the incident energy dependence of the \gls*{RIXS} spectra to extract their spectral weights (integrated intensities). As expected for a Raman-like process in \gls*{RIXS} the peak energies were seen to be independent of incident energy, so this was used as an additional fitting constraint. In Supplementary Fig.~\ref{fig:SI_Edep_magnon}, we can see that the magnon peak intensity has only one maximum near the main resonance peak around $853$~eV. On the contrary, both the double-magnon peak (Supplementary Fig.~\ref{fig:SI_Edep_magnon_zoom_in}) and the exciton peak (Supplementary Fig.~\ref{fig:SI_Edep_exciton}) have two maxima, one near the main resonance peak and the other around $857.7$~eV. Such distinct resonance behaviors have been summarized in Fig.~4\textbf{a}. 

\section{Double-magnon and exciton propagation}
This section provides further analysis of the propagation of the exciton and double-magnon excitations illustrated schematically in Fig.~5 of the main text. For clarity, we will refer to double-magnon propagation via \textit{exchange} and exciton propagation via \textit{hopping}.

\subsection{Double-magnon propagation}
\label{sec:doublemagprop}

\ce{NiPS3} exhibits antiferromagnetic order on a honeycomb lattice with both easy-plane and easy-axis anisotropy in addition to first ($J_1$), second ($J_2$), and third ($J_3$) nearest-neighbor isotropic exchange interactions, with the latter being the largest interaction~\cite{Wildes2022Magnetic, Scheie2023Spin}. Despite these complexities, we can obtain a significant amount of insight by considering a simplified picture, which nonetheless captures the essentials of the interactions at play. Propagation within \ce{NiPS3}'s zig-zag antiferromagnetic ground state can be conceptualized as propagation along either a ferro- or antiferro-magnetic chain direction. We simplify matters further by considering only nearest-neighbor exchange processes in the discussion that follows. (The same procedure applies to longer range exchange processes, as outlined below, so the effects of these terms can also be deduced easily). 

For a ferromagnetic chain, the double-magnon propagation occurs via the exchange of $|1,1\rangle$ and $|1,-1\rangle$, where the states are labeled as $|S,m_S\rangle$ following Fig.~5 of the main text. The amplitude for this process can be calculated using second-order perturbation theory in the spin flip processes, with the intermediate state being nearest-neighbor single magnon states with an energy cost proportional to the sum of the single-ion anisotropy $\propto D$ and the spin exchange $ \propto J_1$. Altogether, we obtain that such a double-magnon can propagate along the ferromagnetic direction at the scale of the order of $\propto J_1^2 / (2D+nJ_1)\sim J_1$, where $n$ is the number of broken magnetic bonds in the intermediate state of perturbation theory. Although such a calculation is in principle only valid in the Ising-like limit, Ref.~\cite{Zheng2005} has shown that it can be extended also to the isotropic case. 

The other propagation direction in the \ce{NiPS3} plane is the antiferromagnetic direction (see Fig.~5 of the main text). For this case, we need to invoke a fourth order process in the spin flip terms with three intermediate states, with the highest energy being equivalent to the cost of having two `extra' double-magnons in a line. This multiplicity is due to the fact that the double-magnon propagation is a composition of two processes along the antiferromagnetic direction, namely 1) the exchange of the $|1,1\rangle$ and $|1,-1\rangle$ states (illustrated in row 2 and 3 of Fig.~5\textbf{b} of the main text), and 2) propagation of a double-magnon via an intermediate state with two `extra' double-magnons on the nearest-neighbor sites. A similar case for the magnon propagation in the antiferromagnet was discussed in Ref.~\cite{Verresen2018quantum}. Due to spin exchanges entering both the denominator and numerator of the perturbative formulae, such a process again leads to an effective propagation at the scale of the spin exchange $J_1$. 

These estimations for the exchange process have implicitly assumed that the double-magnon is a bound state, which is formally justified only in the limit of large Ising anisotropy. For \ce{NiPS3}, the double-magnon probably has high decay rates into two nearest-neighbor single-magnon states. Fortunately, the latter should remain bound (due to attractive interactions between magnons on nearest neighbor sites), so our analysis should remain a good order-of-magnitude estimate. 

We note that the leading $J_3$ exchange process connects only antiferromagnetically aligned Ni spins, which would suppress any difference between propagation along different directions in the lattice. This is in addition to the fact that \ce{NiPS3} is structurally and magnetically twinned, meaning that the ferro- and antiferro-magnetic directions in the lattice are not empirically distinguishable (see the methods section).  Although technically very challenging, the development of ultrahigh energy resolution \gls*{RIXS} under strain could be implemented to study specific magnetic monodomains to more directly test whether the double-magnon is or is not a bound state.

\subsection{Exciton hopping} 
\label{sec:excitonexchange}
Before discussing the amplitude of the exciton hopping, we first consider possible spin exchanges for the $|1,0 \rangle_{\bf i}$ state with $|1,\pm1 \rangle_{\bf j}$ between sites $\bf{i}$ and $\bf{j}$, following the discussion in Ref.~\cite{Takubo2007Unusual}. 
In this case, the relevant terms in the Hamiltonian are 
\begin{equation}\label{eq:Hspin}
\begin{split}
h^\text{spin-flip}&= \frac{J_{{\bf i},{\bf j}}}{2} \left( S^+_{\bf i} S^-_{\bf j} + \text{h.c.} \right) \\
 &= J_{{\bf i},{\bf j}} \big( 
  |1,1\rangle_{\bf j}|1,0\rangle_{\bf i} ~ 
     _{\bf j}\langle 1,0|_{\bf i}\langle 1,1|
    + |1,-1\rangle_{\bf i}|1,0\rangle_{\bf j} ~_{\bf i}\langle 1,0|_{\bf j}\langle 1,-1|
    +\text{h.c.}
 \big), 
 \end{split}
\end{equation} 
where $J_{{\bf i},{\bf j}}$ is the exchange coupling between sites $\bf{i}$ and $\bf{j}$ and $S^\pm_{\bf i}$ are the raising and lowering spin operators on site ${\bf i}$. 

We now turn to the exciton hopping process. Since NiPS$_3$ is a magnetically ordered insulator, an exciton at site ${\bf i}$, denoted by $|0,0 \rangle_{\bf i}$, can only hop via an interchange process with the $S = 1$ states at a neighboring site ${\bf j}$  (denoted here as $|1,\pm 1 \rangle_{\bf j}$). The relevant terms in the Hamiltonian for exciton hopping are similar to Eq.~\eqref{eq:Hspin}, i.e.: 
\begin{align}\label{eq:Hex}
h^\text{hop} = J_{{\bf i},{\bf j}} \big( 
    |1,1\rangle_{\bf i}|0,0\rangle_{\bf j}~_{\bf i}\langle 0,0|_{\bf j}\langle 1,1| 
    + |1,-1\rangle_{\bf i}|0,0\rangle_{\bf j}~_{\bf j}\langle 1,-1|_{\bf i}\langle 0,0|
    + \text{h.c.} \big),
\end{align}
with the prefactors in Eqs.~\eqref{eq:Hspin} and \eqref{eq:Hex} being identical. The underlying reason for this key observation is that the dominant processes involved are (super)exchange processes that take place on the ligand orbitals rather than the nickel orbitals~\cite{Takubo2007Unusual}. As a result, individual $s=1/2$ spin flips cannot happen on the nickel atoms, leading to vanishing amplitudes for terms that would differentiate between the $|0, 0\rangle$ singlet hopping and the $|1, 0\rangle$ triplet exchange.

\subsection{Exciton propagation}
\label{sec:excitonprop}

We start by discussing exciton propagation along the ferromagnetic direction in the zigzag antiferromagnetic ground state. In fact, once the exciton hopping is 
known
[see Eq.~\eqref{eq:Hex} above], this propagation is the easiest to understand: it merely amounts to the free hopping of the exciton in the ferromagnetic background with a hopping amplitude equal to the spin exchange, similar to the double-magnon case.

A more complex situation is encountered for the antiferromagnetic direction. Here exciton propagation is also due to the hopping process described by Eq.~\eqref{eq:Hex} but obtaining a coherent propagation is slightly intricate---as shown in Fig.~5\textbf{a} of the main text. This whole process can essentially be divided into two steps. First, the exciton interchanges twice with the spin background, just as in the ferromagnetic background (second and third row of Fig.~5\textbf{a} of the main text). This leads to the creation of an intermediate state with two `extra' double-magnons situated next to each other. Second, the double-magnons can be annihilated by two spin-exchange processes, in a similar manner as described in \ref{sec:doublemagprop} above. In total, it requires four spin exchanges for the exciton to freely move to the next-nearest-neighbor site, just as the double-magnon does. Such exciton propagation is also at the energy scale proportional to the spin exchange.

Altogether, we observe strong similarities between the way the exciton and the double-magnon move through the spin $S=1$ zigzag antiferromagnetic honeycomb lattice---in particular, both motions require the same number of spin exchanges and the energy scales are in both cases the same and proportional to the spin exchanges. As discussed in the following section, the dominant exciton hopping is through the third nearest neighbors connected by antiferromagnetically aligned spins, therefore, the dispersions along the in-plane $H$ and $K$ directions would be expected to be overall rather similar, just like the double-magnon case. 

\section{Tight-binding model fit to the exciton dispersion}
\label{sec:TBfit}
Tight-binding model approaches have been widely used to model exciton dispersion in molecular solids \cite{Cudazzo2015exciton}. Although the way the Hund's exciton we studied here moves is different to the mechanism for Frenkel excitons (spin exchange processes as discussed above instead of dipole–dipole interactions), tight-binding models can still be useful at a phenomenological level to provide a simple, but informative empirical approach to extract the lengthscale of the effective interactions governing the exciton dispersion.

Using the monoclinic unit-cell notation, we formulated a simplified effective tight-binding model on a two-dimensional honeycomb lattice with isotropic effective ``hopping'' terms $t_n$. We obtain three terms in the exciton dispersion ($E_{t1}$, $E_{t2}$, and $E_{t3}$) associated with first, second, and third nearest neighbor effective interactions ($t_1$, $t_2$ and $t_3$), where

\begin{equation}
\begin{split}
E_{t1} = & t_1 \sqrt[\pm]{4 \cos^2 (\pi H) + 4 \cos (\pi H) \cos (\pi K) + 1} \\
E_{t2} = & 2t_2 \{\cos (2\pi H) + \cos [\pi (H+K)] + \cos [\pi (H-K)] \} \\
E_{t3} = & t_3 \sqrt[\pm]{4 \cos^2 (2\pi H) + 4 \cos (2\pi H) \cos (2\pi K) + 1} .
\end{split}
\end{equation}
By co-fitting the measured low-temperature exciton dispersion along both $H$ and $K$ directions, we obtain the results as shown in Supplementary Fig.~\ref{fig:SI_TBmodel}. For $E_{t1}$ and $E_{t3}$, we selected the sign of the solution based on the empirically observed upward-dispersion trend near the Brillouin zone center. We tested the individual contribution of the three different functional forms and found that the fits with third nearest neighbor effective hopping alone can capture the observed periodicity of the dispersion, indicating that third nearest neighbor interactions play the leading role in the exciton dispersion. As stated before, this phenomenological model is an effective parameterization of a process that fundamentally arises from magnetic exchange and not real hopping. Our observation of leading third nearest neighbor interactions is consistent with the fact the third nearest neighbor spin exchange is dominant in the spin Hamiltonian of \ce{NiPS3} \cite{Scheie2023Spin, Wildes2022Magnetic}. 

\clearpage

\begin{figure*}
\includegraphics{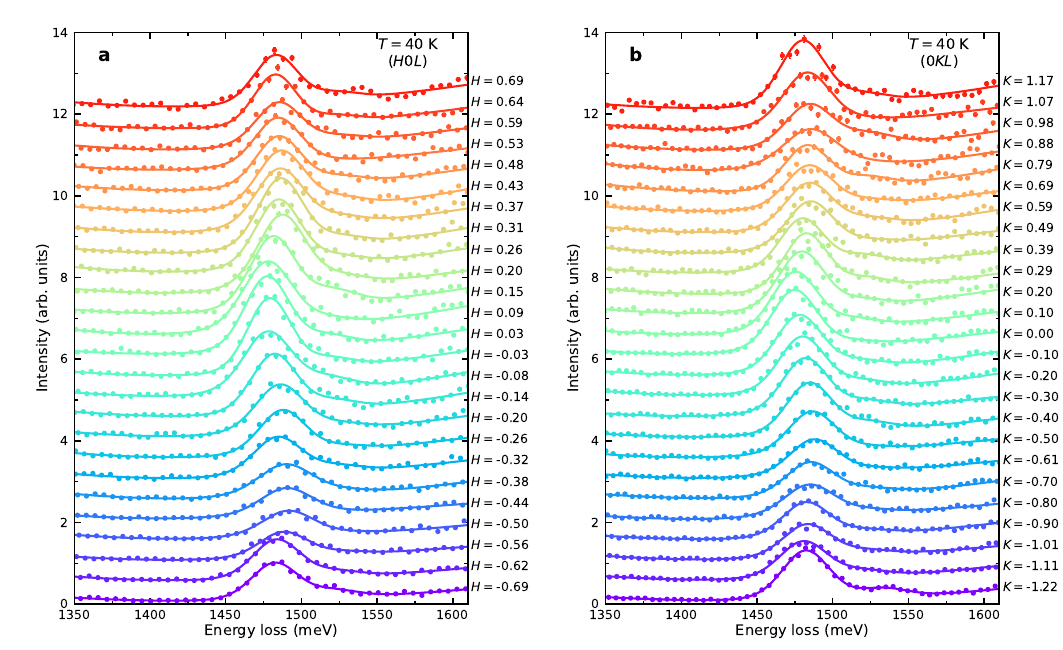}
\caption{\textbf{\gls*{RIXS} spectra measured at $\bm{T=40}$~K with an energy window chosen to isolate the exciton}. \textbf{a}, The exciton peak at various in-plane momentum transfer $H$ measured in the ($H0L$) plane. \textbf{b}, The exciton peak at various in-plane momentum transfer $K$ measured in the ($0KL$) plane. All the measurements were taken with linear horizontal $\pi$ polarization of the incident x-rays at the resonant energy of $853.4$~eV for excitons. The data is the same as the intensity maps shown in Fig.~2\textbf{a} and \textbf{b}. The solid lines are fits to the data with details in Methods Section. Data are shifted vertically for clarity. Error bars represent one standard deviation.}
\label{fig:SI_LT_exciton}
\end{figure*}

\begin{figure*}
\includegraphics{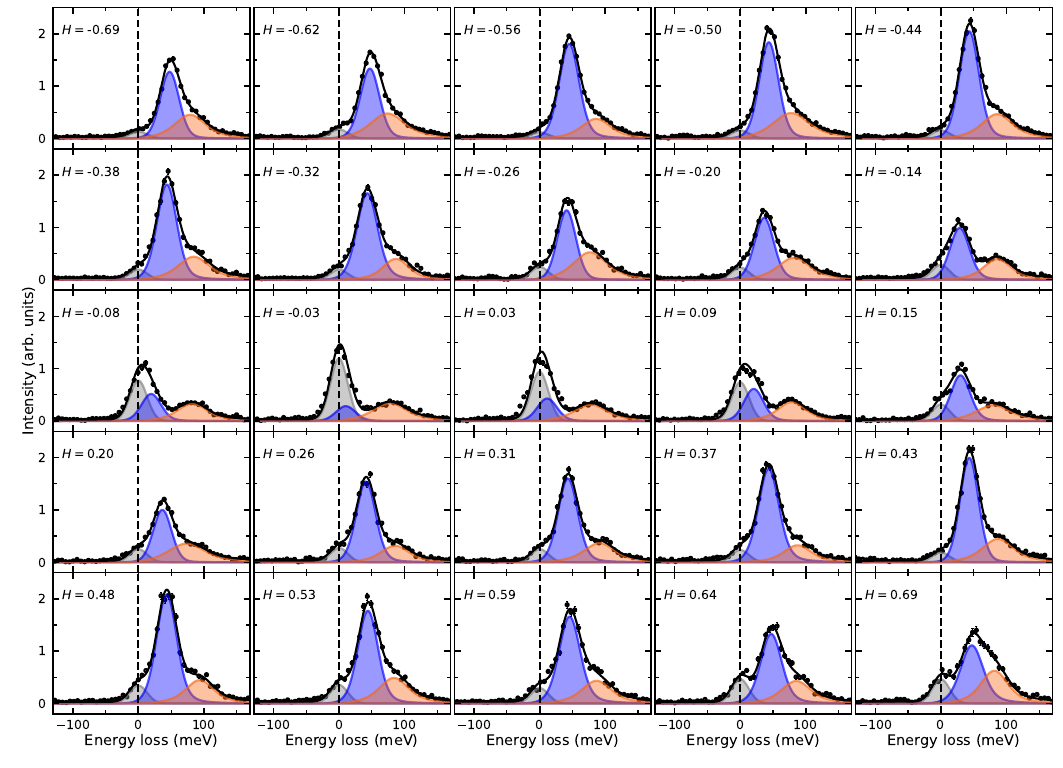}
\caption{\textbf{\gls*{RIXS} spectra measured in the ($\bm{H0L}$) plane at $\bm{T=40}$~K with an energy window chosen to isolate the magnon and double-magnon}. Each panel displays the spectra at a specific in-plane momentum transfer $H$ measured in the ($H0L$) plane. The solid lines are best fits to the data with three components, i.e., the elastic line (gray), the magnon peak (blue), and the double-magnon peak (orange). The detailed description of the fitting can be found in the Methods Section and \ref{sec:RIXS_fits}. The vertical dashed line in each panel labels the energy zero. Error bars represent one standard deviation.}
\label{fig:SI_H0L_LT_magnon}
\end{figure*}

\begin{figure*}
\includegraphics{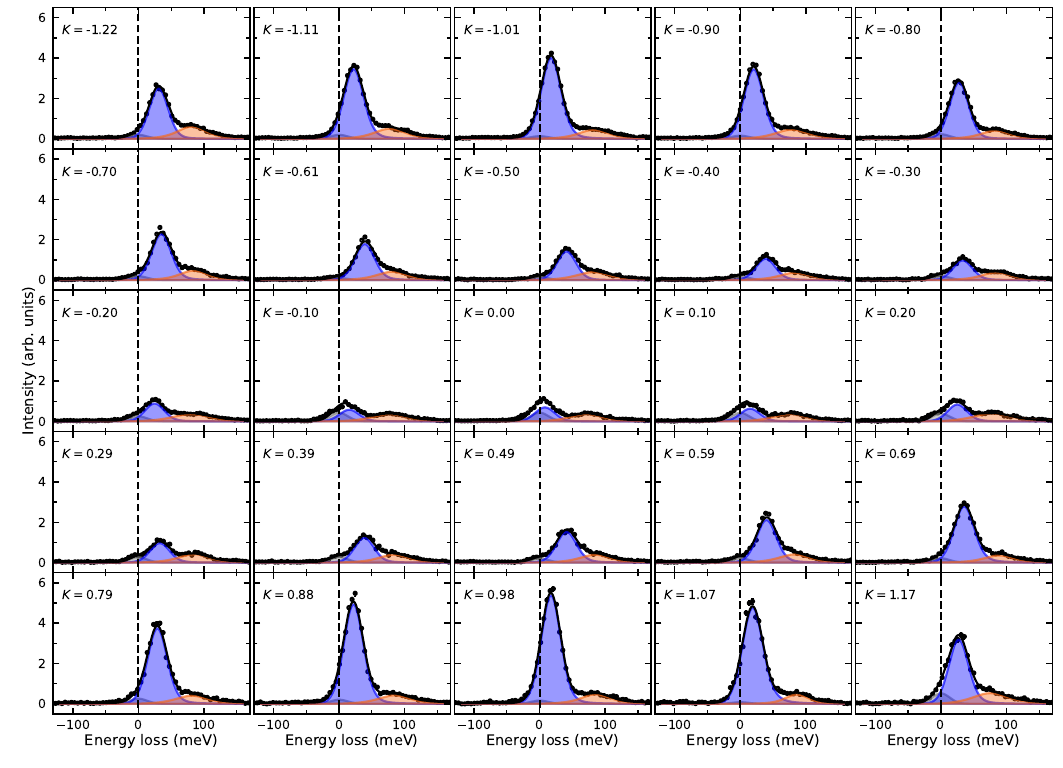}
\caption{\textbf{\gls*{RIXS} spectra measured in the ($\bm{0KL}$) plane at $\bm{T=40}$~K with an energy window chosen to isolate the magnon and double-magnon}. Each panel displays the spectra at a specific in-plane momentum transfer $K$ measured in the ($0KL$) plane. The solid lines are best fits to the data with three components, i.e., the elastic line (gray), the magnon peak (blue), and the double-magnon peak (orange). The detailed description of the fitting can be found in the Methods Section and \ref{sec:RIXS_fits}. The vertical dashed line in each panel labels the energy zero. Error bars represent one standard deviation.}
\label{fig:SI_0KL_LT_magnon}
\end{figure*}

\begin{figure*}
\includegraphics{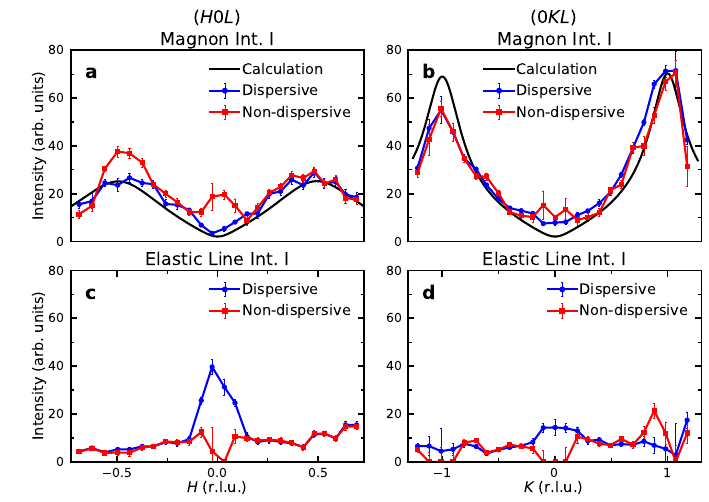}
\caption{\textbf{Fitting result comparison between the best fit (blue circle), which gives a dispersive exciton mode, and the fit assuming a null hypothesis of a non-dispersive exciton (red square) coming from a miscalibration of the spectra}. \textbf{a}--\textbf{b}, Integrated Intensity of the magnon peak measured in ($H0L$) plane \textbf{a} and ($0KL$) plane \textbf{b}, respectively. The black solid lines are the calculated magnon intensities as described in the Methods Section, with the same scaling factor for both panels. \textbf{c}--\textbf{d}, Integrated Intensity of the elastic peak measured in ($H0L$) plane \textbf{c} and ($0KL$) plane \textbf{d}, respectively. We see that the best fit is in good accord with the calculated magnon intensity, whereas the null hypothesis can be excluded as it implies an unphysical decrease in the elastic line intensity at $(0, 0)$, which is the point at which the elastic line intensity would usually be maximized.}
\label{fig:SI_fits_comparison}
\end{figure*}

\begin{figure*}
\includegraphics{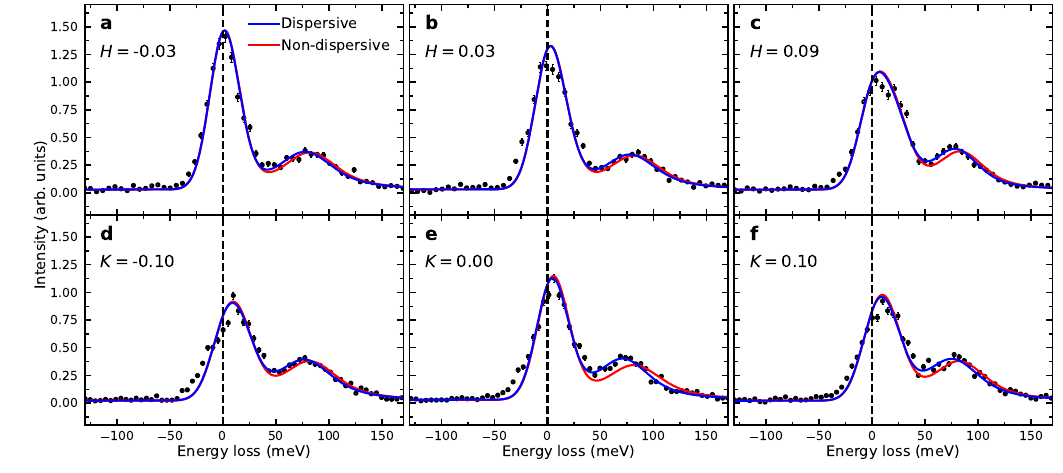}
\caption{\textbf{Representative fitting comparisons between the best fits (blue lines), which give a dispersive double-magnon mode, and the fits assuming a null hypothesis of a non-dispersive double-magnon (red lines)}. \textbf{a}--\textbf{c}, \gls*{RIXS} spectra measured in the ($H0L$) plane at $T=40$~K at three representative in-plane momentum transfers $H$. \textbf{d}--\textbf{f}, \gls*{RIXS} spectra measured in the ($0KL$) plane at $T=40$~K at three representative in-plane momentum transfers $K$. We can clearly see that red lines assuming non-dispersive double-magnons are worse compared to the best fits (blue lines) which give dispersive double-magnons. The vertical dashed line in each panel labels the energy zero. Error bars represent one standard deviation.}
\label{fig:SI_fits_comparison2}
\end{figure*}

\begin{figure*}
\includegraphics{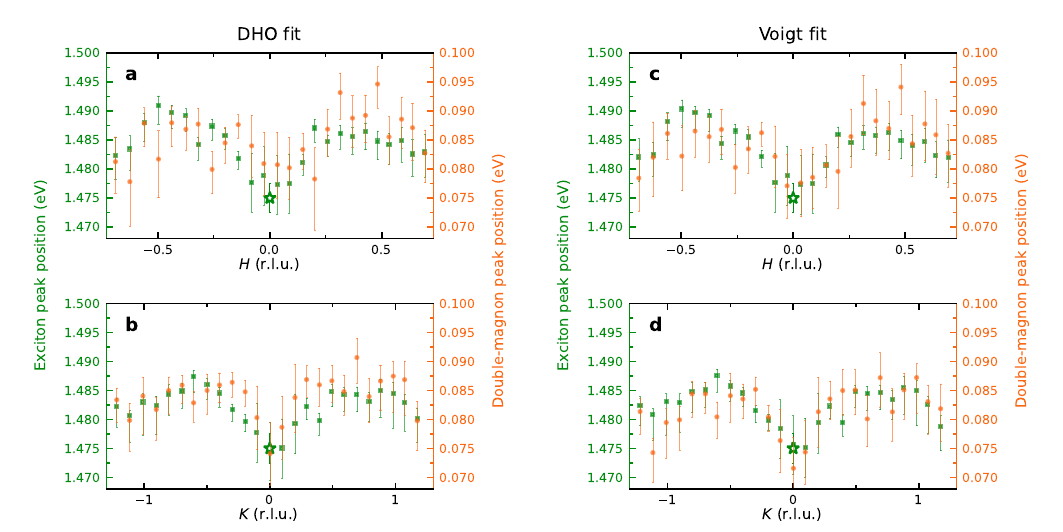}
\caption{\textbf{Fit comparison between the \gls*{DHO} and Voigt models used for the magnon and double-magnon peaks}. \textbf{a}--\textbf{b}, fitted peak positions of the excitons (green squares) and double-magnons (orange circles) using the \gls*{DHO} model for the magnon and double-magnon peaks. The data is the same as the dispersion plots shown in Fig.~2\textbf{e} and \textbf{f}. \textbf{c}--\textbf{d}, fitted peak positions of the excitons (green squares) and double-magnons (orange circles) using the Voigt function for the magnon and double-magnon peaks. The fitting results based on the two different functional forms for the magnons and double-magnons are quite similar, validating the robustness of the exciton and double-magnon dispersions.}
\label{fig:SI_DHO_vs_Voigt}
\end{figure*}

\begin{figure*}
\includegraphics{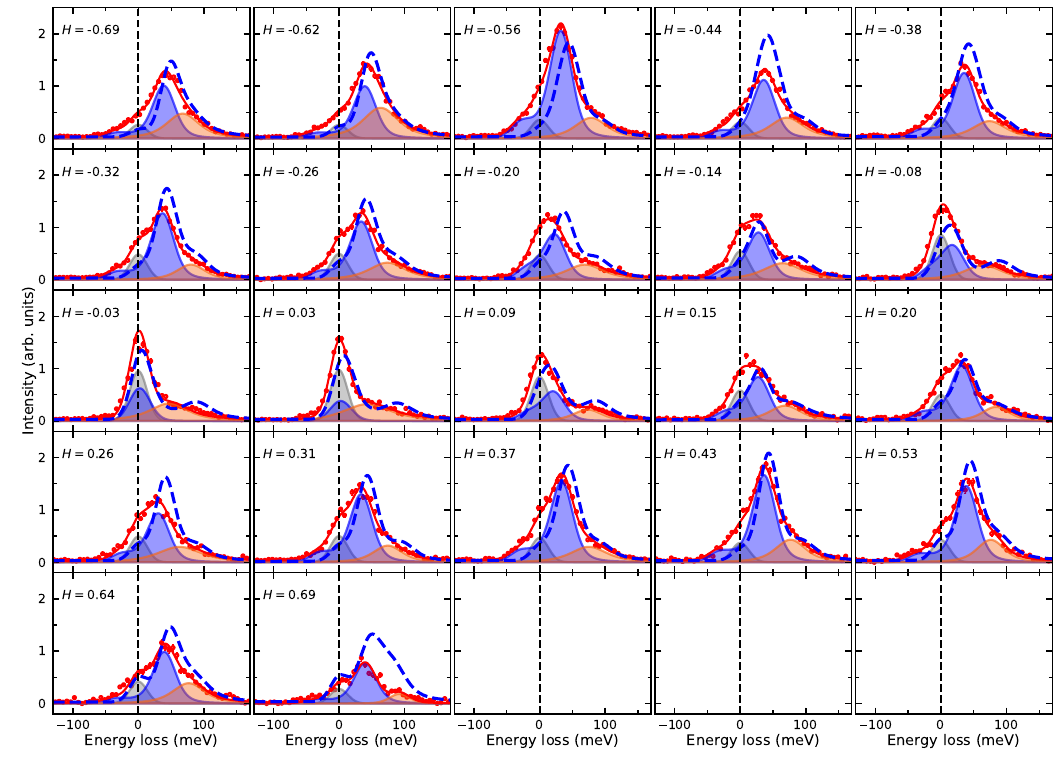}
\caption{\textbf{\gls*{RIXS} spectra measured in the ($\bm{H0L}$) plane at $\bm{T=190}$~K with an energy window chosen to isolate the magnon and double-magnon}. Each panel displays the spectra at a specific in-plane momentum transfer $H$ measured in the ($H0L$) plane at $T=190$~K, above its magnetic ordering temperature. We used linear horizontal $\pi$ polarization of the incident x-rays at the resonant energy of $853.4$~eV for excitons. These data are the same as the intensity maps shown in Fig.~3\textbf{b} and are provided to show the linecuts directly. The solid red lines are best fits to the data with three components, i.e., the elastic line (gray), the magnon peak (blue), and the double-magnon peak (orange). These are provided to clarify the identity of different spectral features. The blue dashed line shows fits to the 40~K data for comparison. The vertical dashed line in each panel labels the energy zero. Error bars represent one standard deviation.}
\label{fig:SI_H0L_HT_magnon}
\end{figure*}

\begin{figure*}
\includegraphics{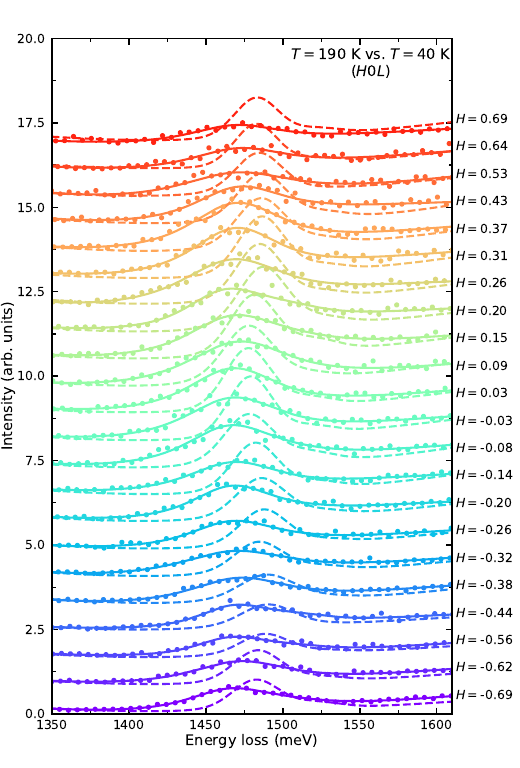}
\caption{\textbf{\gls*{RIXS} spectra measured at $\bm{T=190}$~K with an energy window chosen to isolate the exciton}. The measurements were taken at various in-plane momentum transfer $H$ in the ($H0L$) plane at $T=190$~K, above its magnetic ordering temperature. We used linear horizontal $\pi$ polarization of the incident x-rays at the resonant energy of $853.4$~eV for excitons. These data are the same as the intensity maps shown in Fig.~3\textbf{a} and are provided to show the linecuts directly. The solid lines are fits to the data. The dashed lines show the fitted curves to the corresponding 40~K data for comparison. Data are shifted vertically for clarity. Error bars represent one standard deviation.}
\label{fig:SI_HT_exciton}
\end{figure*}

\begin{figure*}
\includegraphics{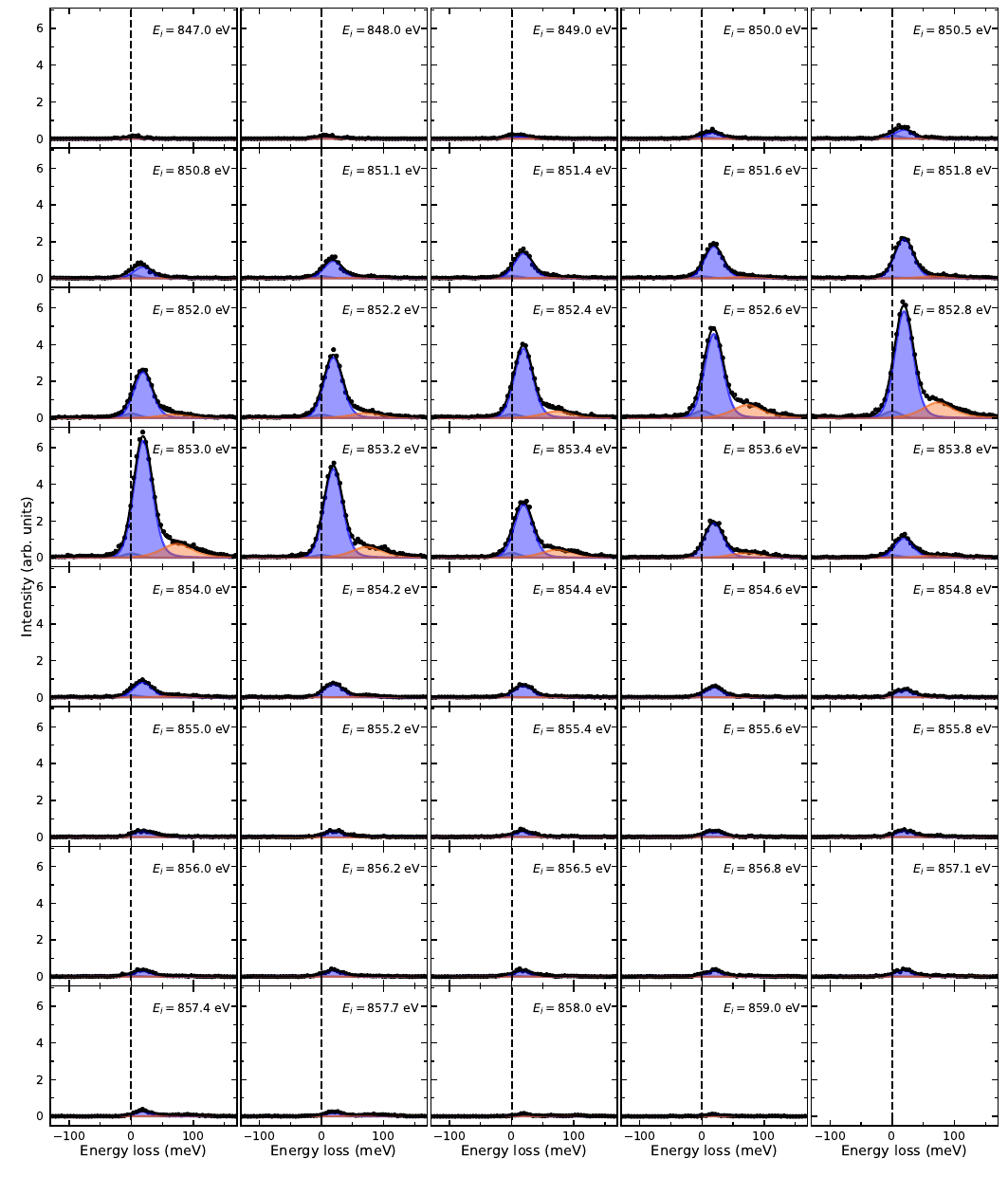}
\caption{\textbf{Incident energy dependence of the \gls*{RIXS} spectra with an energy window chosen to highlight the intensity change of the magnon peak}. Each panel displays the spectra with a specific incident photon energy through the Ni $L_3$ resonance measured in the ($0KL$) plane at $T=40$~K with linear horizontal $\pi$ polarization of the incident x-rays. These data are the same as the intensity maps shown in Fig.~1\textbf{a} and are provided to show the linecuts directly. The solid red lines are best fits to the data with three components, i.e., the elastic line (gray), the magnon peak (blue), and the double-magnon peak (orange). The vertical dashed line in each panel labels the energy zero. Error bars represent one standard deviation.}
\label{fig:SI_Edep_magnon}
\end{figure*}

\begin{figure*}
\includegraphics{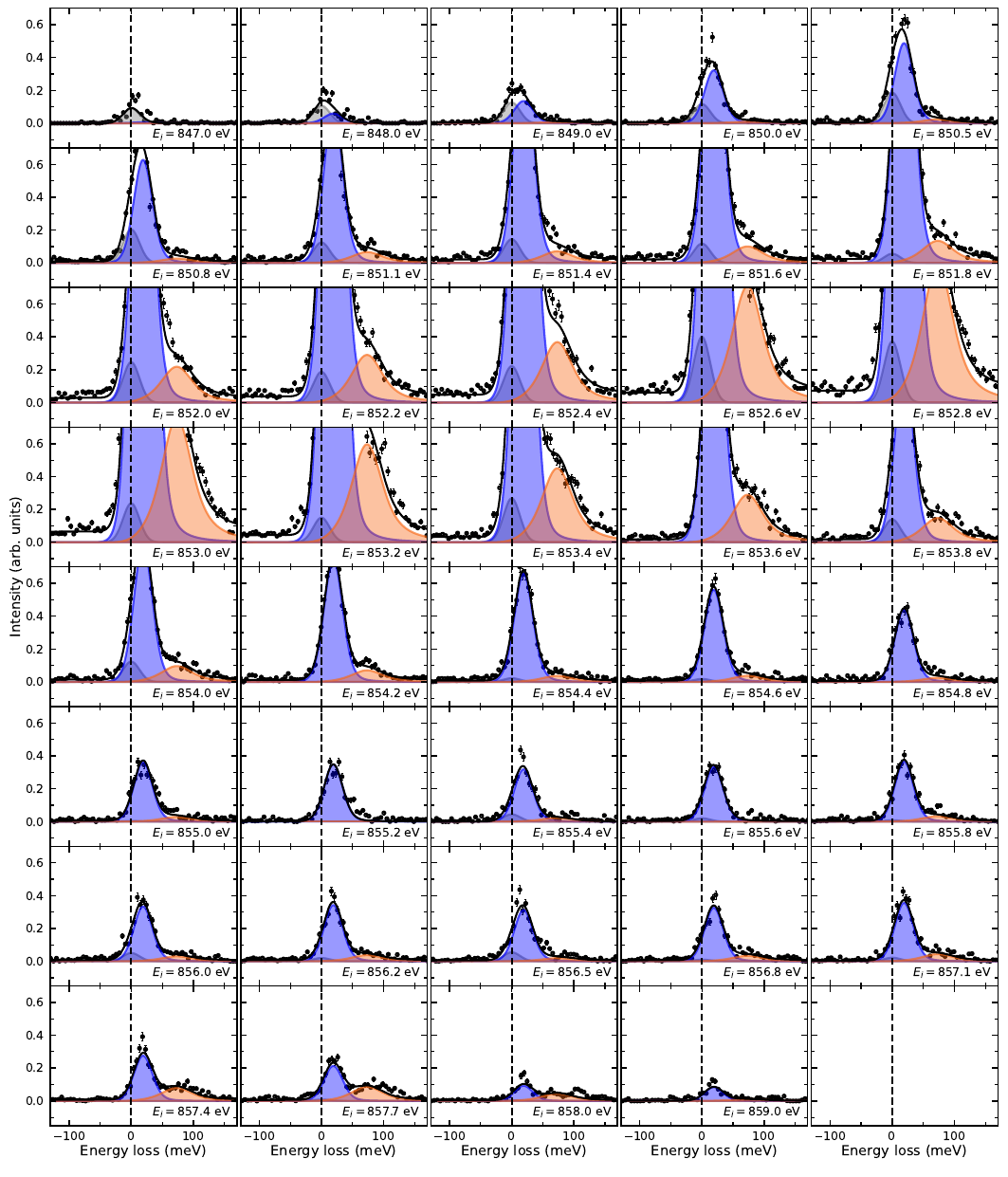}
\caption{\textbf{Incident energy dependence of the \gls*{RIXS} spectra with an energy window chosen to highlight the intensity change of the double-magnon peak}. Each panel displays the spectra with a specific incident photon energy through the Ni $L_3$ resonance measured in the ($0KL$) plane at $T=40$~K with linear horizontal $\pi$ polarization of the incident x-rays. These data are the same as that shown in Supplementary Fig.~\ref{fig:SI_Edep_magnon} but zoomed in on the double-magnon peak (orange). Error bars represent one standard deviation.}
\label{fig:SI_Edep_magnon_zoom_in}
\end{figure*}

\begin{figure}
\includegraphics{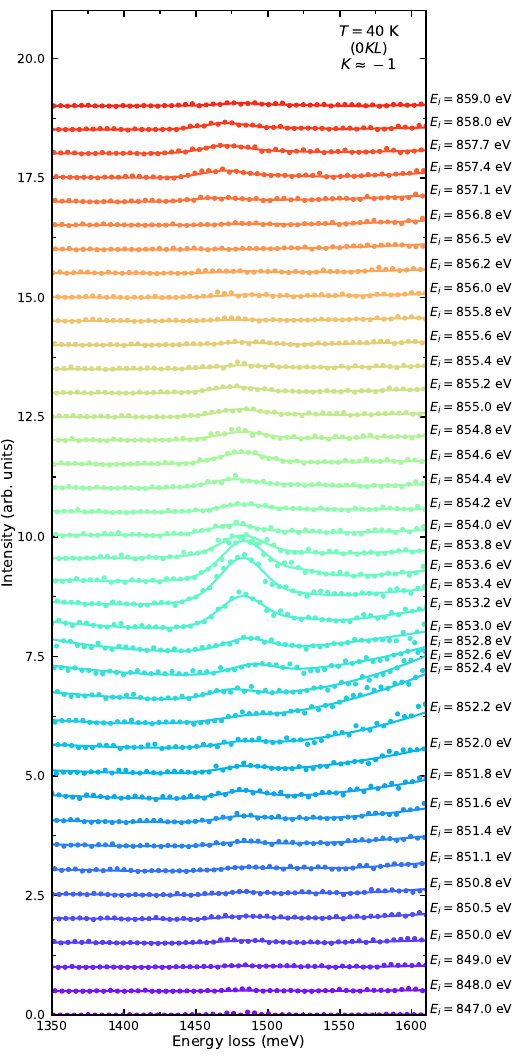}
\caption{\textbf{Incident energy dependence of the \gls*{RIXS} spectra with an energy window chosen to highlight the intensity change of the exciton peak}. The measurements were taken at various incident photon energy through the Ni $L_3$ resonance measured in the ($0KL$) plane at $T=40$~K with linear horizontal $\pi$ polarization of the incident x-rays. These data are the same as the intensity maps shown in Fig.~1\textbf{a} and are provided to show the linecuts directly. The solid lines are the fits to the data with details in Methods Section. Data are shifted vertically for clarity. Error bars represent one standard deviation.}
\label{fig:SI_Edep_exciton}
\end{figure}

\begin{figure}
\includegraphics{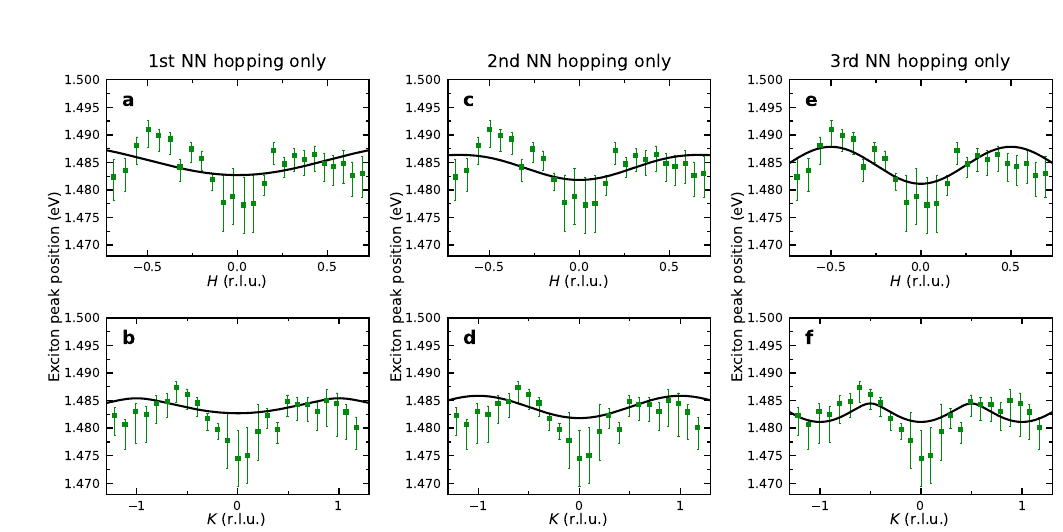}
\caption{\textbf{Tight-binding model fits to the exciton dispersion}. Each column displays tight-binding model fits with only first (\textbf{a}--\textbf{b}), second (\textbf{c}--\textbf{d}), and third (\textbf{e}--\textbf{f}) nearest neighbor interactions, respectively. The green squares are the measured exciton dispersion as a function of the $H$ and $K$ in-plane momentum transfer, respectively. The black lines are fitted curves. The best fitted values are $t_1=1.4(5)$~meV, $t_2=-0.5(1)$~meV, and $t_3=1.7(3)$~meV, respectively. Error bars represent one standard deviation.}
\label{fig:SI_TBmodel}
\end{figure}

\clearpage

\bibliographystyle{naturemag}
\bibliography{refs}